%% file: main.tex
\documentclass[aps,prx,10pt,twocolumn,amsmath,amssymb,superscriptaddress, nofootinbib]{revtex4-2}
\usepackage{graphicx} 
\usepackage{mathtools}
\mathtoolsset{showonlyrefs}
\usepackage{bm} 
\usepackage{hyperref}
\usepackage{braket,bbold,color}
\usepackage{comment}

\usepackage[left=2cm,right=2cm,top=2cm,bottom=3cm]{geometry}
\usepackage{graphicx}
\usepackage{amsfonts}
\usepackage{amsmath}
\usepackage{mathtools}
\usepackage{comment}

\usepackage[english]{babel}

\usepackage{enumerate} 
\usepackage{nicefrac}
\DeclareUnicodeCharacter{0308}{}

\usepackage{amsthm}
\usepackage{float}
\usepackage[font={small},figurename={FIG.}]{caption}
\usepackage{stackengine}
\usepackage{dsfont}
\usepackage{algpseudocode}
\usepackage{algorithm}
\usepackage{xr}
\usepackage{zref}
\usepackage{mathtools}
\usepackage[export]{adjustbox}
\usepackage{enumitem}
\usepackage[x11names,dvipsnames]{xcolor}
\usepackage{kbordermatrix}
\usepackage{tikz}
\usetikzlibrary{matrix}

\DeclareMathAlphabet{\mymathbb}{U}{BOONDOX-ds}{m}{n}

\DeclarePairedDelimiter\floor{\lfloor}{\rfloor}

\makeatletter
\DeclareRobustCommand{\cev}[1]{%
  \mathpalette\do@cev{#1}%
}
\newcommand{\do@cev}[2]{%
  \fix@cev{#1}{+}%
  \reflectbox{$\m@th#1\vec{\reflectbox{$\fix@cev{#1}{-}\m@th#1#2\fix@cev{#1}{+}$}}$}%
  \fix@cev{#1}{-}%
}
\newcommand{\fix@cev}[2]{%
  \ifx#1\displaystyle
    \mkern#23mu
  \else
    \ifx#1\textstyle
      \mkern#23mu
    \else
      \ifx#1\scriptstyle
        \mkern#22mu
      \else
        \mkern#22mu
      \fi
    \fi
  \fi
}

\theoremstyle{definition}

\theoremstyle{definition}

\newtheorem{theorem}{Theorem}

\usepackage{amsmath}

\newcommand{\bs}[1]{\boldsymbol{#1}}

\newcommand{\Trace}[1]{\text{Tr}\left(#1\right)}

\newcommand\arrowline{\mathrel{\ooalign{$\rightarrow$\cr%
  \kern-.15ex\raise.275ex\hbox{\scalebox{1}[0.522]{$\mid$}}\cr}}}

\newcommand{\mc}[1]{\mathcal{#1}}

\newcommand{\Rank}{\text{rank}}
\newcommand{\supp}{\text{supp}}

\newcommand{\pnk}{\node[fill=pink]{ };}
\usepackage{soul}

\begin{document}
\input{seq1}

\end{document}

%% file: seq1.tex
\title{Contextuality in Sequential State Discrimination}

\author{Nyan Raess}
\affiliation{Royal Holloway, University of London}
\author{Farid Shahandeh}
\affiliation{Royal Holloway, University of London}

\begin{abstract}
Generalized contextuality is known to be required in optimal strategies for quantum state discrimination protocols. More recently, sequential discrimination tasks have been studied; $n$ players attempt to determine in which state a qubit was prepared, in such a way that they all have a finite probability of success. We consider the extent to which contextuality plays a role in sequential versions of both unambiguous and minimum error discrimination. In the standard nonsequential case where $n=1$, we use the COPE formalism to demonstrate that the presence of contextuality is guaranteed not only for the optimal measurement, but for a specific set of nonoptimal measurements as well. In the sequential case $n>1$, we show that the presence of contextuality depends on which states are prepared, and on the protocol (unambiguous or minimum error) selected. 
\end{abstract}

\maketitle
\setlength{\parskip}{0pt}
\section{Introduction} \label{sec: intro}

State discrimination is a key task in quantum information processing. 
It comes in a variety of forms, such as minimum error state discrimination (MESD), unambiguous state discrimination (USD), and maximum confidence discrimination~\cite{Croke2006}.
All forms of state discrimination can be phrased as one-way communication games between two players, Alice and Bob, that go as follows.
Alice sends a random message $i$ encoded in  a quantum state $\rho_i$ from a set $\mc{S}=\{\rho_i\}_i$ and sends it to Bob.
Bob performs a measurement aimed at  identifying the message $i$.
However, the figure of merit varies across different variations.
In MESD, Bob's goal is to minimize the chance of misidentifying the state. 
For MESD of two quantum states, the best strategy is well-known, and is called the \textit{Helstrom} measurement~\cite{Helstrom}. 
In USD, on the other hand, 
Bob performs a measurement that \textit{never} misidentifies the state~\cite{Ivanovic1987,Dieks1988,Peres1988}.
The price he pays is that some results will be deemed inconclusive and must be discarded. 
The optimal measurement for the case of two states  with arbitrary prior probabilities was found by Jaeger and Shimony~\cite{Jaeger1995}, and a complete geometric picture, including an analytic treatment of the three-state case, was given in Ref.~\cite{Bergou2012}. For a comprehensive review of state discrimination, see Ref.~\cite{Bae2015}.

Contextuality, on the other hand, has been widely studied as a resource for quantum advantage in computation \cite{Bermejo-Vega2017, Wallman2012, Frembs2018, Lillystone2018} and many other information-processing tasks~\cite{multiplex,Pan2025,Ambainis2016,Gupta2022,galvao}. 
In the original form introduced by Kochen and Specker \cite{KS67, Specker60}, contextuality refers to the impossibility of assigning fixed values to all quantum observables simultaneously.
This was later extended to generalized contextuality by relaxing the outcome determinism assumption~\cite{Spekkens2005},
defining it as the impossibility of assigning fixed probability distributions and response functions to all quantum states and measurement effects, respectively.
It was then shown to be present in MESD~\cite{Schmid2018} and in USD and maximum confidence discrimination~\cite{Flatt2022}. More specifically, the optimal strategies in each case cannot be explained by a noncontextual (NC) ontological model. Recently, it has been shown that \textit{conclusive exclusion} also exhibits generalized contextuality~\cite{conc_exc}.
Here, Bob's goal is to \textit{rule out} a state index $i$ with certainty.

In this paper, we characterize the generalized contextuality of sequential state discrimination. In the sequential setting, Alice encodes her message in a quantum state and passes it to Bob, who performs a discriminating measurement and passes his post-measurement state to the next player, Charlie.
This cascades until Alice's message reaches the $n$th player~\cite{Bergou2013,Fields2020,Zhang2018,seq-mesd}.
We determine the extent to which contextuality continues to play a role as the number of players increases.  We will find that in both cases this depends heavily on the confusability of the states. Surprisingly, contextuality is more likely to appear in sequential USD when the states have low confusability, whereas in MESD the opposite is true.

Our paper is organized as follows.
In Sec.~\ref{sec:SD} we describe USD and MESD in the case of two pure qubit states, and, following \cite{Bergou2013,seq-mesd}, give the optimal solutions in both the single-player and sequential cases.
In Sec.~\ref{sec:GC} we review the basics of generalized contextuality~\cite{Spekkens2005} and the recent linear-algebraic approach via the \textit{conditional outcome probabilities of events} (COPE) matrix of Refs.~\cite{shahandeh2024cloning-1,rank-sep}.
This allows us to recast the results of \cite{Flatt2022} and draw a parallel between contextual advantages in MESD and USD for the 1-player case. We present our results for the sequential case in Sec.~\ref{sec:results}, including a comparison of the relative prevalence of contextuality in sequential USD as compared to MESD, followed by a more detailed discussion of the difference between the protocols. Discussions and conclusions are presented in Sec.~\ref{sec:concl}.

\section{State Discrimination} \label{sec:SD}

\subsection{Unambiguous State Discrimination}

\subsubsection{Single-player case}

In the single-player USD, Alice samples a quantum
state $\rho_i$  from a set $\mc{S}=\{\rho_1, \rho_2\dots \rho_n\}$ according to a prior distribution $\{p_i\}_{i=1}^n$.
The set $\mc{S}$ and the prior probabilities are known to Bob.
His objective is to determine which state (i.e. the index $i$) he has received while ensuring that the probability of error is zero.
That is, he must perform a measurement in which  the outcome $i$ \textit{cannot} occur if the state is $\rho_j, j\neq i$. 
Clearly, this is only possible if each density operator $\rho_i$ has a kernel $\mc{K}_i\in \mc{H}$ such that $\mc{K}_i\nsubseteq \mc{K}_j$ $\forall \ i,j$.
These yield
a POVM $\{\Pi_0,\Pi_1, \Pi_2,\dots \Pi_n\}$, where for all $i>0$, $\Pi_i\in \bigcap_{j\ne i}  \mc{K}_{j}$ so that $\Trace{\Pi_i\rho_j} >0$  if and only if $i=j$. 
In the nontrivial case, where at least some states are mutually nonorthogonal, the price he pays for zero error is the outcome $\Pi_0$, necessary for the completeness $\sum_{i=0}^{n}\Pi_i=\mathds{1}$. 
Thus, $\Pi_0$ 
corresponds to the inconclusive outcome, where Bob cannot be sure which state he has received.  
He therefore seeks to minimize the probability  of the inconclusive outcome,
\begin{equation} \label{eq:USD_gen_fail_p}
    \sum_{i=1}^n p_i\Trace{\Pi_0\rho_i}.
\end{equation}
%

Let us consider the simplest case of a single qubit prepared in one of two pure states $\mc{S}=\{\ket{\psi_1}\bra{\psi_1},\ket{\psi_2}\bra{\psi_2}\}$ with equal prior probabilities $p_1=p_2=1/2$.
Note that mixed states of a qubit do not have kernels, implying that a
USD of mixed qubit states is impossible.
Furthermore, the case  with arbitrary priors is addressed in Ref.~\cite{Jaeger1995}.
Bob's POVM reduces to
$\{\Pi_1^B, \Pi_2^B, \Pi_0^B\}$ 
where
\begin{equation}
\Pi_1^B = c_1^B\ket{\psi_2^\perp}\bra{\psi_2^\perp}, \quad \Pi_2^B = c_2^B\ket{\psi_1^\perp}\bra{\psi_1^\perp}, 
\label{eq:pi12form}
\end{equation}
and
\begin{equation}
\Pi_0^B=\mathds{1}-c_1^B\ket{\psi_2^\perp}\bra{\psi_2^\perp}-c_2^B\ket{\psi_1^\perp}\bra{\psi_1^\perp}.
\label{eq:pi0form}
\end{equation}
Here, $\ket{\psi_i^\perp}$ is the state orthogonal to $\ket{\psi_i}$, and $c_i^B \in [0,1]$ are the tuneable parameters available to Bob. If the first outcome occurs, Bob can be sure that the state was $\ket{\psi_1}$, and similarly for $\ket{\psi_2}$. 
As mentioned, whenever the states are non-orthogonal there must be an inconclusive outcome $\Pi_0^B$ , and whenever it occurs, Bob has failed.

Bob's performance in discriminating his
inputs is quantified by  $s^2=|\braket{\psi_1|\psi_2}|^2$, called the \textit{confusability}. The higher the confusability, the less likely he is to distinguish one state from the other. 
Since only two states are being discriminated, we may assume that $s$ is real and positive~\cite{Bergou2013,Fields2020,seq-mesd}. 
Therefore, we write:
\begin{equation}
\begin{split}
    \ket{\psi_2}=s\ket{\psi_1}+\sqrt{1-s^2}\ket{\psi_1^\perp} \\
    \ket{\psi_2^\perp}=\sqrt{1-s^2}\ket{\psi_1}-s\ket{\psi_1^\perp}.
    \label{eq:psi2_psi1}
    \end{split}
\end{equation}
These reduce Bob's failure, as given by Eq.~\eqref{eq:USD_gen_fail_p}, to
\begin{equation}
\begin{split}
    Q_1^B &= \frac{1}{2}(q_1^B+q_2^B) \\
    &=1-\frac{c_1^B+c_2^B}{2}(1-s^2) \label{eq:q-failure},
    \end{split}
\end{equation}
where  $q_i^B=\bra{\psi_i}\Pi_0^B\ket{\psi_i}=1-c_i^B(1-s^2)$.

\begin{figure}
\centering
    \includegraphics[width=0.8\columnwidth]{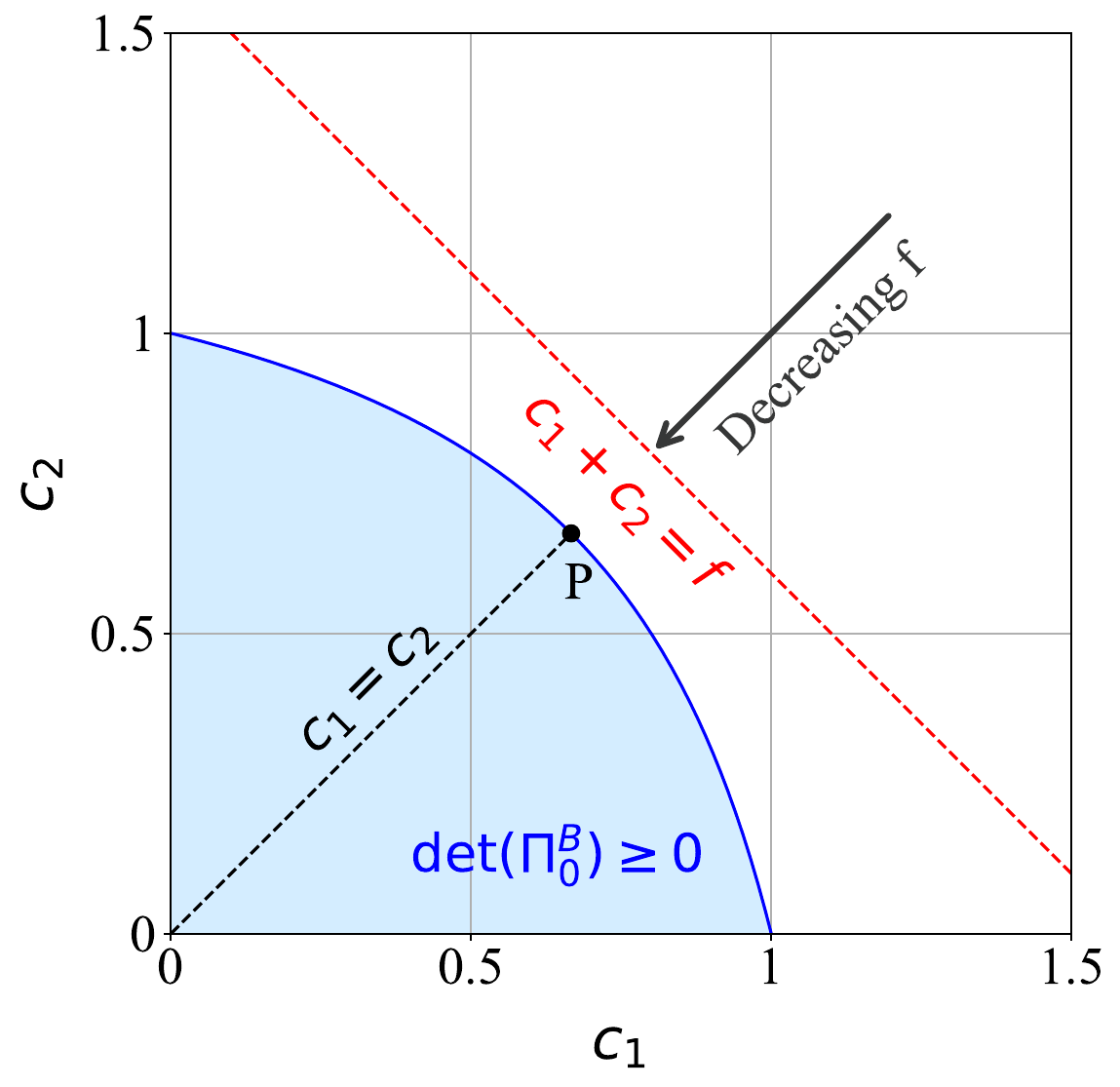}
    \caption{A visualization of the optimal solution for USD of two pure qubit states under equal priors, here with confusability $s^2=1/4$. The blue region contains all feasible $c_1^B,c_2^B$ choices i.e. $\Pi_0^B\geq 0$. To minimize the failure probability we want to maximize the sum $c_1^B+c_2^B$: we want the largest $f$ so that the red line intersects the feasible region. It is easy to see that this will occur along the line $c_1^B=c_2^B$; specifically, at $P = \left(\nicefrac{1}{1+s}, \nicefrac{1}{1+s}\right)$.}
    \label{fig:c1c2feasible}
\end{figure}

Clearly, not all choices of $c_i^B$ lead to a valid measurement: while the operators $ \Pi_1^B, \Pi_2^B$ are positive for $c_i^B\in [0,1]$, $\Pi_0^B$ becomes negative if the $c_i^B$ are too large. 
 Writing $\Pi_0^B$ in the $\{\ket{\psi_1},\ket{\psi_1^\perp}\}$ basis,
\begin{equation}
    \Pi_0^B = \begin{bmatrix}
        1-c_1^B(1-s^2) & c_1^Bs\sqrt{1-s^2} \\
        c_1^Bs\sqrt{1-s^2} & 1-c_1^Bs^2-c_2^B
        \label{eq:pi0_matrix}
    \end{bmatrix},
\end{equation}
reduces Bob's task  to choosing $c_1^B$ and $c_2^B$ such that $Q_1^B$ is minimized, while
\begin{equation}
    \det(\Pi_0^B)=1-c_1^B-c_2^B+c_1^Bc_2^B(1-s^2) \geq 0.
    \label{eq:det}
\end{equation}
This is equivalent to
\begin{equation}
\begin{split}
    &\max_{c_i^B} \ c_1^B+c_2^B \notag \\
    &\text{s.t.} \ \ \det{\Pi_0^B}\geq0.
    \label{eq:USD-task}
    \end{split}
\end{equation}
Given that
\begin{equation}
    \frac{\partial \det{\Pi_0^B}}{\partial c_i^B} = -1+c_j^B(1-s^2) \leq 0,
    \label{eq:det_dep}
\end{equation}
the optimal solution must occur when the determinant is minimized.  
When the determinant is zero, i.e., $\Pi_0^B$ is rank-one,
\begin{equation}
c_2^B = \frac{c_1^B-1}{c_1^B(1-s^2)-1},
\label{eq:0detalt}
\end{equation}
and maximizing $c_1^B+c_2^B$, we find the well-known solution \cite{book,disc_rev}:
\begin{flalign}
   &c_1^B=c_2^B=\frac{1}{1+s}:=c_{\rm{opt}}, \\
   &Q_1^B = s
   \label{eq:opt_q}
\end{flalign}
Figure~\ref{fig:c1c2feasible} depicts the optimal solution above.
We will later see that the analogous statement does not hold for the NC model.

\subsubsection{Sequential USD} \label{sec:seq_USD}

In the sequential case, Alice  randomly samples one of the two states $\mc{S}=\{\ket{\psi_1}\bra{\psi_1},\ket{\psi_2}\bra{\psi_2}\}$ and passes it onto Bob, who performs an unambiguous measurement of the form given in Eqs.~\eqref{eq:pi12form} and~\eqref{eq:pi0form}. Upon receiving the state $\ket{\psi_i}$, he either fails with probability $q_i^B$ or succeeds with probability $1-q_i^B$, and  passes his post-measurement state to Charlie. 
Charlie then performs his unambiguous discrimination and passes on the state to the next player.
Assuming that, including Alice, there are $n+1$ players, $n$ of which make a measurement, this continues until the $(n+1)$-th player makes the final measurement. 
The rules of the game forbid the communication of each player's measurement outcome to the next.

However, just as Alice's set of states $\mc{S}:=\mc{S}_1$ is known to Bob, the  $j$th player's set of post-measurement states, denoted by $\mc{S}_{j}$,   is known to the next player. 
The goal of the game is for each player to independently determine Alice's message, i.e., the index $i$.

It is well-known that post-measurement states are not uniquely defined by a POVM \cite{Bergou2004rev,Bae2015}. This means that the protocol is fully specified only when each player has selected their post-measurement states. With this in mind, define $\Pi_i^B=A_i^\dagger A_i$ where $A_i$ are Bob's detection operators. These operators can be used to define Bob's post-measurement states. Their nonuniqueness is immediate  since any choice of the form $A_i=U_i(\Pi_i^B)^{1/2}$ is valid for arbitrary unitary $U_i$.  In the most general case:
\begin{equation}
    \begin{split}
    &A_1=\sqrt{c_1^B}\ket{\phi_1}\bra{\psi_2^\perp} \notag\\
    &A_2=\sqrt{c_2^B}\ket{\phi_2}\bra{\psi_1^\perp} \notag\\ A_0=&\sqrt{a_1^B}\ket{\chi_1}\bra{\psi_2^\perp}+\sqrt{a_2^B}\ket{\chi_2}\bra{\psi_1^\perp},
    \label{eq:detection_Bob}
    \end{split}
\end{equation}
where $a_i^B$ and $\ket{\chi_i}$ are to be determined. 
Since Charlie does not know whether Bob  succeeds or fails, in general, he receives a mixed state of the form\begin{equation}
\ket{\psi_i'}\bra{\psi_i'}= q_i^B\ket{\chi_i}\bra{\chi_i}+(1-q_i^B)\ket{\phi_i}\bra{\phi_i}.
\end{equation}
Recall, however, that for a single qubit USD of mixed states is impossible. This mandates the choice $\ket{\chi_i}=\ket{\phi_i}$. That is, if Bob receives $\ket{\psi_i}$, his post-measurement state will be $\ket{\phi_i}$ \textit{regardless} of whether he succeeds or fails. Thus, Charlie's task is to unambiguously discriminate $\{\ket{\phi_1}\bra{\phi_1},\ket{\phi_2}\bra{\phi_2}\}$.

We will focus on the case $n=2$, where Charlie is the last player. Our results easily generalize to anarbitrary number of players. Using $\ket{\phi_i}=\ket{\chi_i}$, the completeness relation $\sum_{i=0}^2A_i^\dagger A_i=\mathds{1}$  implies that
\begin{equation}
a_i=\frac{q_i^B}{1-s^2}=\frac{1}{1-s^2}-c_i^B.
\end{equation}
Defining Charlie's confusability as $t^2:=|\braket{\phi_1|\phi_2}|^2$,  $\Pi_0^B=A_0^\dagger A_0$ yields the consistency relation
\begin{flalign}
    q_1^Bq_2^B&=\left[1-c_1^B(1-s^2)\right]\left[1-c_2^B(1-s^2)\right] \notag \\
    &=s^2/t^2 ; \label{eq:cnstcy_1}
\end{flalign}

Any $c_1^B,c_2^B\in [0,1]$ satisfying this will implement a measurement that discriminates two states with confusability $s^2$, and whose post-measurement states have confusability $t^2$, with $s\leq t\leq 1$. Specifying the value $t$ therefore replaces the nonnegativity of the determinant, Eq.~\eqref{eq:det}, with the equality constraint Eq.~\eqref{eq:cnstcy_1}.

It is useful to compare these two constraints. Adding $c_1^Bc_2^B(1-s^2)$ to both sides of Eq.~\eqref{eq:cnstcy_1} yields
\begin{equation}
    \begin{split}
        \det(\Pi_0^B)&=s^2\left[1/t^2-c_1^B-c_2^B+c_1^Bc_2^B(1-s^2)\right],\label{eq:det_const} \\ &= s^2(\det{\Pi_0^B+1/t^2-1}).
    \end{split}
\end{equation}
When the determinant is set to 0, which is necessary for Bob's optimal measurement, we find\begin{equation}
\frac{1}{t^2}-1=0.
\end{equation}
Therefore, when Bob's measurement effects are rank-one operators, the only possible post-measurement confusability is $t^2=1$, \textit{regardless} of whether his choice is optimal or not. 

It follows that for Charlie to receive distinguishable states, i.e. $t<1$, Bob's effect $\Pi_0^B$ must satisfy $\det(\Pi_0^B)>0$. To determine Bob's optimal measurement for fixed $s$ and $t$, we note that Bob is minimizing the failure probability $Q^B_1=\frac{1}{2}(q_1^B+q_2^B)$ subject to the constraint $q_1^Bq_2^B=s^2/t^2$. By the AM-GM inequality, $Q^B_1\ge q^B_1q^B_2$, with equality if and only if $q^1_B=q^2_B$, which in turn requires
\begin{equation}
    c_1^B=c_2^B=\frac{1-s/t}{1-s^2}
    \label{eq:c12-branch-seq}
\end{equation}
For fixed $s$ and $t$, this choice of $c_1^B$ and $c_2^B$ therefore gives the minimal failure probability for Bob,
\begin{equation}
    Q_1^B=\frac{s}{t}.
    \label{eq:s-on-t-fail}
\end{equation}
As Charlie is the last player, there is no need for his post-measurement states to be distinguishable. This means that he is free to make an optimal measurement on his input states. Following Eq.~\eqref{eq:opt_q} he will choose $c_1^C=c_2^C=1/(1+t)$ which results in the minimal failure probability, $t$. 
Importantly, the probability of both Bob and Charlie failing given by  $Q_1^BQ_1^C = (s/t) \cdot t =s$ does not depend on $t$ \cite{Bergou2013}.

With Bob's choices of $c_1^B$ and $c_2^B$ for a given $s$ and any desired $t$ at hand, it remains only to determine the optimal value of $t$ such that Bob and Charlie's joint probability of success  given by \begin{equation}
(1-Q_1^B)(1-Q_1^C) = (1-s/t)(1-t).
\end{equation}
is maximized.
We thus find that $t=\sqrt s$  is the optimal choice. 
The generalization to $n+1$ players similarly fixes the ratio of pre- and post-measurement confusabilities. In particular if the $j$-th player receives states $\ket{\psi_i^{(j)}}$ with \begin{equation}\braket{\psi_1^{(j)}|\psi_2^{(j)}}=s_j,\end{equation} then we have \cite{Bergou2013},
\begin{flalign}
    s_1=s, \ \frac{s_j}{s_{j+1}}= s^{1/n} \implies s_j=s^{(n-j+1)/n}.
    \label{eq:obs_overlaps}
\end{flalign}
This yields minimum failure probability $s^{1/n}$ for \textit{each} player and so the probability that all players successfully identify the state is \cite{Bergou2013}:
\begin{equation}
    P^{(q)}_{\mathrm{USD}}(n)= \left(1-s^{1/n}\right)^n
    \label{eq:opt_n_obs}
\end{equation}
Clearly, $\lim_{n\rightarrow\infty}P^{(q)}_{\mathrm{USD}}=0$.
However, the probability that all players fail is given by $s$, which is independent of the number of players.

Finally, note that asking Bob and Charlie to perform an optimal measurement with \textit{fixed} pre- and post-measurement confusabilities singles out the choice $c_1=c_2$ for all players. As noted in \cite{Pang2013}, \textit{directly} optimizing the joint success probability of all players  $\Pi_{j=1}^{n}(1-Q^{(j)}_1)$ under the condition that all operators are positive leads to the so-called boundary solutions, where for a certain range in $s$, the choice $c_1=0$ and $c_2=1$, or the reverse, becomes optimal for all players. The POVM becomes $\{\ket{\psi_j}\bra{\psi_j},\ket{\psi_j^\perp}\bra{\psi_j^\perp}\}$ and there is no way to unambiguously identify the state $\ket{\psi_i}$. Since this constitutes unambiguous discrimination of only one state, we follow Ref.~\cite{Bergou2013} and choose to ignore such strategies.

\subsection{Minimum Error State Discrimination}

\subsubsection{Single-player case}

As mentioned in the introduction, the difference between MESD and USD scenarios is Bob's objective.
In MESD Alice samples one of the states $\{\ket{\psi_1},\ket{\psi_2}\}$ at random and sends it to Bob,  whose task is to correctly identify the state with the highest possible probability. Unlike in the USD, he is allowed to make errors, so there is no need to designate a POVM element for an inconclusive outcome.  

Bob's POVM now consists of two elements $\{M_1^B, M_2^B\}$, $M_1^B+M_2^B=\mathds{1}$, and if he observes outcome $M_i$ he guesses that the state index was $i$. The probability $p_{bi}$ that he successfully identifies $i$ is given by
\begin{equation}
p_{bi}=\bra{\psi_i}M_i^B\ket{\psi_i}.
    \label{eq:MESD_meas}
\end{equation}
The goal is to maximize the probability of correctly identifying either state, i.e., $P_1^B=(p_{b1}+p_{b2})/2$.
We are free to assume that the states still satisfy Eq.~\eqref{eq:psi2_psi1}.
The optimal choice is the well-known Helstrom measurement:
\begin{flalign}
    M_1^B&= \ket{g_{\psi_1}}\bra{g_{\psi_1}}, M_2^B = \ket{g_{\psi_1}^\perp}\bra{g_{\psi_1}^\perp}, \notag \\
    \ket{g_{\psi_1}}&=\frac{\left(1+\sqrt{1-s^2}\right)\ket{\psi_1}-s\ket{\psi_1^\perp}}{\sqrt{2(1+\sqrt{1-s^2})}},
    \label{eq:helstrom_form}
\end{flalign}
which achieves the success probability
\begin{equation}
\label{eq:helstrom_prob}
    P_1^B=\frac{1}{2}(p_{b1}+p_{b2})=\frac{1}{2}\left(1+\sqrt{1-s^2}\right)
\end{equation}
with $p_{b1}=p_{b2}$.

\subsubsection{Sequential MESD}

The protocol for the sequential MESD closely follows that of sequential USD.  Similar to the sequential USD case, we assume, without loss of generality, that there are only three players: Alice, Bob, and Charlie. We discuss the generalization to $n+1$ players at the end of this section.

Since each player must send distinguishable states to the next, it is again convenient to define detection operators $A_i$ such that $M_i^B=A_i^\dagger A_i$ and 
\begin{flalign}
A_1&=\beta_{11}\ket{v_{11}}\bra{\psi_2^\perp}+\beta_{12}\ket{v_{12}}\bra{\psi_1^\perp} \notag \\
A_2&=\beta_{21}\ket{v_{21}}\bra{\psi_2^\perp}+\beta_{22}\ket{v_{22}}\bra{\psi_1^\perp}
\label{eq:MESD_Det-ops},
\end{flalign}
which encode the post-measurement states~\cite{seq-mesd}. 
Recall that, in the sequential USD,  Bob's post-measurement state depended only on the state he received and not on his  success or failure. Although this is not strictly necessary in MESD, we will nonetheless impose that the post-measurement states of each player carry no information about their outcomes. 
This amounts to the choice $\ket{v_{11}}=\ket{v_{21}}:=\ket{v_1}, \ket{v_{12}}=\ket{v_{22}}:=\ket{v_2}$ in Eq.~\eqref{eq:MESD_Det-ops}, so that $\ket{v_i}$ is \textit{always} Bob's  post-measurement state, if Alice sent $\ket{\psi_i}$.

Using Eq.~\eqref{eq:MESD_meas}, we thus find,
\begin{equation}
    s/t=\sqrt{p_{b1}(1-p_{b2})}+\sqrt{p_{b2}(1-p_{b1})}
    \label{eq:st_const_MESD}
\end{equation}
where $t:=\braket{v_1|v_2}$ is the post-measurement overlap.
If Bob receives states with a confusability $s^2$ and wishes to produce outputs with confusability $t^2$, his measurement parameters $p_{bi}$ must satisfy Eq.~\eqref{eq:st_const_MESD}. The POVM  effects parametrized with $s,t$ and $p_{bi}$ are:
\begin{widetext}
    \begin{equation}
        M_1^B =
{\renewcommand{\arraystretch}{1.6}%
\begin{bmatrix}
    p_{b1} &
    \dfrac{t\sqrt{p_{b1}(1-p_{b2})}-p_{b1}s}{\sqrt{1-s^2}} \\[6pt]
    \dfrac{t\sqrt{p_{b1}(1-p_{b2})}-p_{b1}s}{\sqrt{1-s^2}} &
    \dfrac{p_{b1}s^2+1-p_{b2}-2st\sqrt{p_{b1}(1-p_{b2})}}{1-s^2}
\end{bmatrix}}, \ M_2^B=\mathds{1}-M_1^B
\label{eq:ME-meas}
\end{equation}
\end{widetext}
Setting the determinant to 0 yields $t=1$, so rank-one strategies are ruled out, as was the case in the sequential USD protocol.

\begin{figure}
\centering
    \includegraphics[width=0.8\columnwidth]{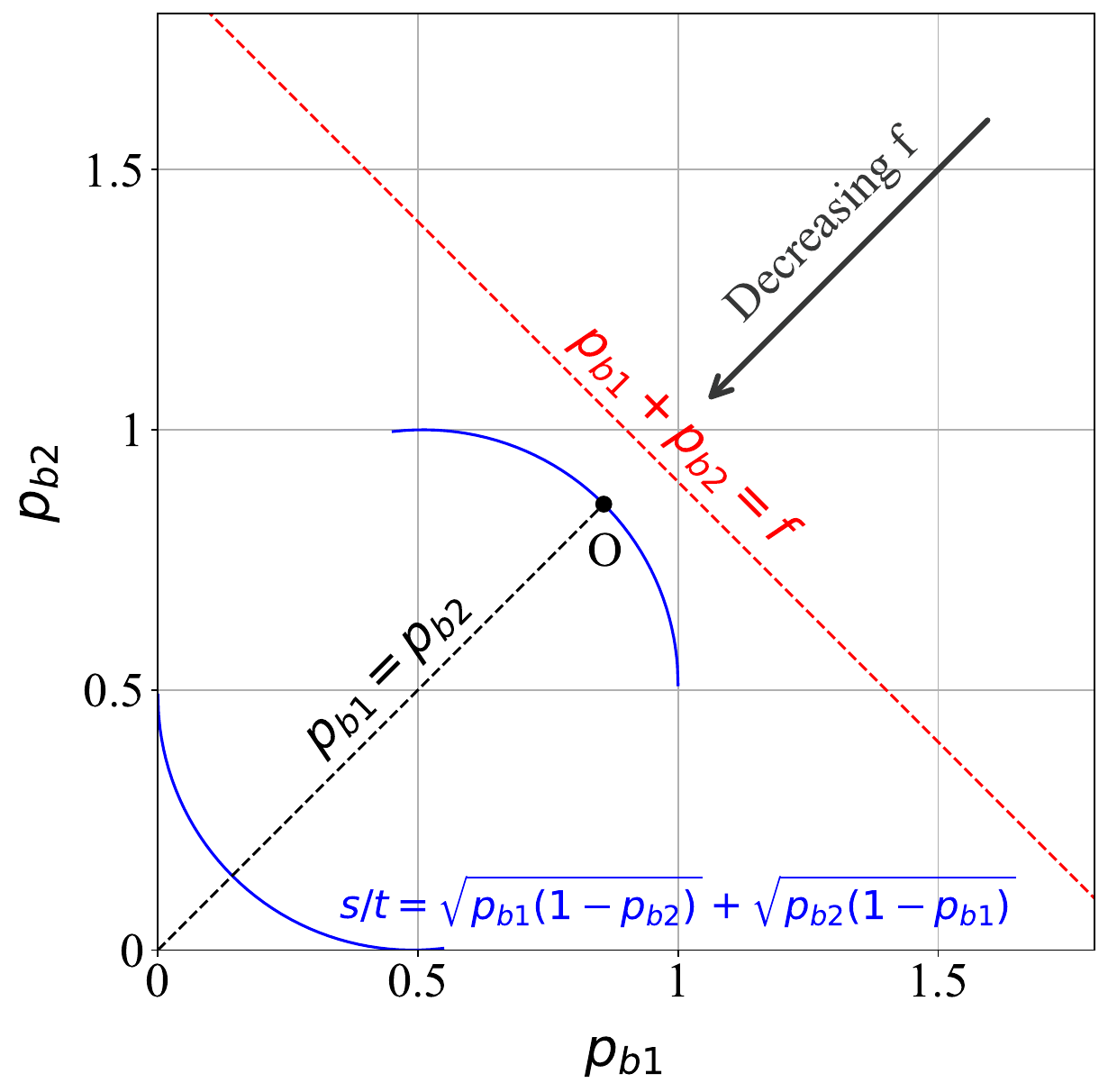}
    \caption{A visualization of the optimal solution for sequential MESD of two pure qubit states under equal priors, assuming the ratio $s/t$ is fixed; here $s/t=0.7$. The blue curve contains all feasible $p_{1b},p_{2b}$ choices. The success probability to be maximized is $1/2(p_{1b}+p_{2b})=f/2$. The largest $f$ for which the red line and blue curve intersect occurs along $p_{1b}=p_{2b}$, specifically at $O = \left(1/2(1+\sqrt{1-s^2/t^2}), 1/2(1+\sqrt{1-s^2/t^2})\right)$.}
    \label{fig:p1p2feasible}
\end{figure}

The sequential MESD also shares some other features of the sequential USD. For instance, the constraint in Eq.~\eqref{eq:st_const_MESD}  singles out $p_{b1}=p_{b2}$ as Bob's optimal choice  for fixed $s$ and $t$, as shown in Fig.~\ref{fig:p1p2feasible}. 
We now calculate the optimal $t$.
Charlie, as the last player, is free to make an optimal (Helstrom) measurement. Under these conditions, their joint success probability is 
\begin{equation}
    \frac{1}{2}(p_{b1}p_{c1}+p_{b2}p_{c2}) = \frac{1}{4}\left(1+\sqrt{1-s^2/t^2}\right)\left(1+\sqrt{1-t^2}\right),
\end{equation}
which is maximized at $t=\sqrt{s}$. Now the generalization to $n+1$ players is immediate: The confusability of post-measurement states follows the sequence defined in Eq.~\eqref{eq:obs_overlaps} and we find \cite{seq-mesd},
\begin{equation}
    P_{\mathrm{MESD}}^{(q)}(n)=\left[\frac{1}{2}\left(1+\sqrt{1-s^{2/n}}\right)\right]^n
    \label{eq:opt-n-obs-mesd}
\end{equation}
Although the strategy achieving this success rate is always available, it is not guaranteed to be optimal for all values of $s$. For a given $n$ it is guaranteed to be optimal only below some critical value $s_b(n)$ which must be numerically calculated \cite{seq-mesd}. As in sequential USD, we restrict our analysis to this strategy as it is the optimal strategy in which every player is equally likely to distinguish either state; for $s>s_b(n)$, the optimal strategy favours one state over the other. 

\section{Generalized Contextuality}\label{sec:GC}

\subsection{Framework}

Generalized contextuality is a key nonclassical feature of quantum theory~\cite{Spekkens2005}. %
It refers to the fact that any ontological model compatible with the statistics of quantum systems cannot assign a unique description to all statistically indistinguishable  states and effects in the theory. In particular, we consider the statistics of a so-called prepare-and-measure (PM) scenario, in which a single system that is prepared in a number of ways, and undergoes a number of measurements.  Let $\mathcal{P} = \{P_1,P_2 \dots\}$ be the set of possible preparation instructions of the system, and $\mathcal{M}=\{M_1, M_2, \dots\}$ the set of possible measurement recipes. Denoting each measurement outcome  by $k$,  the PM scenario is specified by the set of probabilities $p(k|P,M)$ for all possible combinations of $k,P,M$. 

Now suppose that there is an ontological description of these statistics. That is, there exists some underlying `ontic' space $\Lambda$ that completely describes the system's properties. Then each preparation procedure $P$ induces a distribution $\mu_P(\lambda)$ over $\Lambda$, representing the probability that, when performing the instructions $P$, the prepared ontic state is $\lambda$. Each $\mu_P$ is called an \textit{epistemic state} (ES). Similarly,  each measurement-outcome pair  $(M,k)$  corresponds to some nonnegative function $\xi_M^k(\lambda)$ that captures the probability that a measurement $M$ yields outcome $k$, given that the ontic state is $\lambda$. Each $\xi_M^k(\lambda)$ is called a \textit{response function} (RF). As each preparation must place the system in some ontic state, and each measurement must have some outcome, we have,
\begin{equation}
    \sum_{\lambda\in\Lambda}\mu_P(\lambda)=1 \ \forall \ P, \ \ \sum_k\xi_M^k(\lambda)=1 \ \forall \ \lambda, M.
    \label{eq:ont-mod-normalis}
\end{equation}
The probabilities in the PM scenario are given by:
\begin{equation}
    p(k|P,M)=\braket{\xi_M^k, \mu_P}:=\sum_{\lambda\in\Lambda}\mu_P(\lambda)\xi_M^k(\lambda).
    \label{eq:bayes}
\end{equation}
In this way the ontic space $\Lambda$ accounts for all the statistics.
The above construction is called an \textit{ontological model} for the PM scenario \cite{Spekkens2005}. 


Generalized contextuality is a statement about the representation of operationally equivalent procedures in the ontological model.  
Two preparations $P, P'$ are operationally equivalent,  written as $P\sim P'$, if and only if,
\begin{equation} \label{eq:equiv_prep}
   p(k|P,M)=p(k|P',M) \ \forall \  (M,k) 
\end{equation}
This means that two operationally equivalent preparations cannot be separated by any measurement procedure.
Similarly, two measurement outcomes $(M,k)$ and $(M',k')$ are operationally equivalent if and only if they cannot be separated through any preparation procedure. This is denoted by $(M,k)\sim (M',k')$, if and only if
\begin{equation} \label{eq:equiv_eff}
    p(k|P,M)=p(k'|P,M') \ \forall P 
\end{equation}
%

Spekkens' insight was that an NC model should satisfy:
\begin{equation}
\begin{split}
    &P \sim P' \iff \mu_P=\mu_{P'} \\ &(M,k) \sim (M',k') \iff \xi_M^k=\xi_{M'}^{k'} \ .
    \end{split}
\end{equation}
That is, operationally indistinguishable procedures should have a single description at the ontological level. This is a natural assumption to make; without it, the statistics from any experiment would not necessarily tell us anything about the underlying reality. Scenarios for which no NC ontological model can be devised are called \textit{contextual}.

The COPE (conditional outcome probabilities of events) formalism is a linear-algebraic formulation  of generalized contextuality \cite{rank-sep}. We first note that only effects, i.e. outcome-measurement pairs, appear in the statement of generalized contextuality, never measurements and outcomes separately. For this reason, it suffices to consider a joint measurement, in which we flip an unbiased coin and pick a measurement procedure at random. This uniformly rescales the associated probabilities, but, as shown in Ref.~\cite{rank-sep}, in no way affects the presence or absence of contextuality. We now have probabilities $p(k|P_i)$, where it is understood that the index $k$ runs over all the outcomes of every measurement. We collect  these  in a COPE matrix,  
\begin{equation}
C=
    \begin{pmatrix}
        p(1|P_1) & p(1|P_2) & \cdots & \\
        p(2|P_1) & p(2|P_2) & \cdots & \\
        \vdots & \vdots & \ddots
    \end{pmatrix}
    \label{eq:COPE}.
\end{equation}
Via Eq.~\eqref{eq:bayes}, an ontological model is some nonnegative matrix factorization $C=RE$, where $R$ is the matrix of RFs and $E$ the matrix of ESs. The inner dimension of this factorization is $|\Lambda|$. The joint requirements that 1) the ontological model is a linear model of  the operational theory, and 2) operationally equivalent procedures have a unique ontological representation,  result in the \textit{equirank condition}~\cite{rank-sep},
\begin{equation}
    \text{rank}(C)=\text{rank}(R)=\text{rank}(E).
    \label{eq:rank}
\end{equation}
This guarantees that the description provided by $R$ and $E$ of the statistics in $C$ contains no redundancy. 

A nonnegative factorization satisfying Eq.~\eqref{eq:rank} is referred to as an ENMF (\textit{equirank nonnegative matrix factorization)}. Therefore, any COPE matrix which fails to admit an ENMF corresponds to a \textit{contextual} operational theory. In general, however, we have access only to some subset of measurements and states, forming some \textit{fragment} of the COPE of the whole theory. Here, we focus on fragments comprised of states and measurements relevant to our discrimination protocols.


\subsection{Rank guarantees optimal single-player contextuality}
\label{sec:rank-based}

Before moving to the sequential case, we use the COPE formalism to certify that
contextuality powers the optimal single-player MESD and USD protocols. This can be shown using only the rank of the associated COPE matrices and the zeroes present in them.
Let us form a square $6\times 6$ fragment COPE matrix $F$. For USD, the columns correspond to the quantum states
\begin{equation*}
\begin{split}
        &\big\{\ket{\psi_1}\bra{\psi_1}, \ket{\psi_1^\perp}\bra{\psi_1^\perp}, \ket{\psi_2}\bra{\psi_2}, \ket{\psi_2^\perp}\bra{\psi_2^\perp},\\
        &\ket{\pi_0}\bra{\pi_0}, \ket{\tilde{\pi}_0}\bra{\tilde{\pi}_0}\big\},
\end{split}
\end{equation*}
and the rows correspond to the quantum effects,
\begin{equation*}
\begin{split}
        &\big\{\tilde{\Pi}_1^B/2, \tilde{\Pi}_2^B/2, \tilde{\Pi}_0^B/2, \Pi_1^B/2, \Pi_2^B/2, \Pi_0^B\big/2\}.
\end{split}
\end{equation*}
In the above, $\{\Pi_1^B/2, \Pi_2^B/2, \Pi_0^B/2\}$ is the POVM given in Eqs.~\eqref{eq:pi12form} and~\eqref{eq:pi0form} with the optimal rank-one inconclusive effect $\Pi_0^B$ as per Eq.~\eqref{eq:0detalt}.
Furthermore, $\{\tilde{\Pi}_1^B, \tilde{\Pi}_2^B, \tilde{\Pi}_0^B\}$ is the measurement corresponding to the complementary USD game played with the set of states $\{\ket{\psi_1^\perp}\bra{\psi_1^\perp},\ket{\psi_2^\perp}\bra{\psi_2^\perp}\}$.
Lastly, $\ket{\pi_0}$ and $\ket{\tilde{\pi}_0}$ are vectors in the kernel of $\Pi_0^B$ and $\tilde{\Pi}_0^B$, respectively. 

For MESD, both rows and columns correspond to:
\begin{equation*}
\begin{split}
        &\big\{\ket{\psi_1}\bra{\psi_1},\ket{\psi_1^\perp}\bra{\psi_1^\perp},\ket{\psi_2}\bra{\psi_2},\\ &\ket{\psi_2^\perp}\bra{\psi_2^\perp},\ket{g_{\psi_1}}\bra{g_{\psi_1}},\ket{g_{\psi_1}^\perp}\bra{g_{\psi_1}^\perp}\},
\end{split}
\end{equation*}
where $\ket{g_{\psi_1}}$ and $\ket{g_{\psi_1}^\perp}$ are given by~\eqref{eq:helstrom_form}.

%
In either case, the fragment COPE matrix has the structure,
\begin{center}
    \begin{equation}
        D = \vcenter{\hbox{%
        \begin{tikzpicture}
        \tikzset{square matrix/.style={
            matrix of nodes,
            column sep=-\pgflinewidth, row sep=-\pgflinewidth,
            nodes={draw,
              minimum height=#1,
              anchor=center,
              text width=#1,
              align=center,
              inner sep=0pt
            },
          },
          square matrix/.default=1.2cm
        }
        \matrix[square matrix = 0.5cm, left delimiter=(, right delimiter = )]
        {
        \pnk & 0 & \pnk & \pnk & \pnk & \pnk \\
        0 & \pnk &\pnk & \pnk & \pnk & \pnk \\
        \pnk & \pnk  & \pnk & 0 & \pnk & \pnk \\
        \pnk & \pnk & 0 & \pnk & \pnk & \pnk \\
        \pnk & \pnk & \pnk & \pnk & \pnk & 0 \\
        \pnk & \pnk & \pnk & \pnk & 0 & \pnk \\
        };
        \end{tikzpicture}}}
        \label{eq:COPE_struc}
    \end{equation}
\end{center}
where all highlighted cells are nonzero as long as Bob chooses $c_i^B \neq 0$ for USD or $p_{bi}>0$ for MESD. Each adjacent pair of columns or rows now represents an orthogonal pair of vectors in the Bloch sphere. Therefore, the three pairs of adjacent columns (rows) sum to the vector $\bs{1}$ (or proportional to $ \bs{1}^\top$), where $\bs 1$ is the vector of all ones. Moreover, the rank of $D$ is always 3. This is guaranteed by the fact that the statistics are generated by coplanar vectors in the Bloch sphere. 

In Appendix~\ref{sec:app0}, we prove that any rank 3 nonnegative matrix sharing the structure of $D$ admits no ENMF. To claim contextuality as the resource for advantage in state discrimination, we leverage two theorems from Ref.~\cite{Doosti-cloning} to lift the contextuality from the fragment to the full operational theory of a qubit.
\begin{theorem}\cite{Doosti-cloning}\label{thm4}
    Suppose $C_{\rm F}$ is a fragment COPE matrix of another fragment COPE matrix $C'_{\rm F}$.
Suppose $\Rank{C_{\rm F}}=\Rank{C'_{\rm F}}=r$ and that $C_{\rm F}$ does not admit a nonnegative matrix factorization satisfying the equirank condition~\eqref{eq:rank}.
Then, $C'_{\rm F}$ does not admit a nonnegative matrix factorization satisfying the equirank condition.
\end{theorem}
\noindent Assume $C'_{\rm F}=D$ in Theorem~\ref{thm4}. This is a fragment COPE matrix of the fragment operational theory $\rm F$ defined by all states and measurements in the Bloch plane in which these states and effects live. 
It follows that, as $D$ does not admit an ENMF, neither does the subtheory $\rm F$.
Finally,
\begin{theorem}\cite{Doosti-cloning}\label{thm3}
    Suppose $C_{\rm F}$ is a fragment of a COPE matrix $C$ and satisfies relative tomographic completeness.
Then, if $C_{\rm F}$ does not admit an ENMF, neither does $C$.
\end{theorem}
\noindent Note that the fragment $\rm F$, containing all states and effects in a Bloch plane, is relatively tomographically complete.
This means that if there are two identical rows or columns in $C_{\rm F}$, then there is no state or measurement, respectively, in the parent operational theory--in our case, the qubit theory--that can separate the two rows or columns.   
Hence, we can conclude from the violation of Eq.~\eqref{eq:rank} in the USD/MESD task that the full qubit theory does not admit an ontological model satisfying Eq.~\eqref{eq:rank}, and the qubit theory is contextual.
Accordingly, the statement that ``the USD/MESD is contextual'' rests on the line of reasoning developed above, which establishes the contextuality of qubit theory as a necessary resource for the statistical performance of the quantum USD/MESD protocol. 

Note that this proof relies only on two properties of the COPE matrix, namely its zeroes and its rank. Accordingly, any three orthogonal pairs of Bloch vectors lying in the same plane are guaranteed to produce contextual statistics, provided that both the set of preparations and the set of measurements includes all three pairs or rescalings of them.





\subsection{Noncontextual Success Bounds}

In the sequential setting, measurements consisting of rank-one effects are forbidden for any player who is \textit{not} at the end of the chain, because they leave no opportunity for subsequent players to extract Alice's message. On the other hand, a measurement consisting of effects which are not rank-one cannot produce the zeroes in Eq.~\eqref{eq:COPE_struc}, so the results of the previous section cannot be applied directly to the sequential case. We now use the COPE formalism to determine the maximum success rates for USD and MESD which admit an NC explanation. \\

We begin with an operational interpretation of the confusability of two preparations \cite{Schmid2018}. This is relevant as, for both USD and MESD, the success probability is determined by the confusability
$s^2 = |\braket{\psi_1|\psi_2}|^2$ of the two states. \\

Suppose we perform the projective measurement
$\{\ket{\psi_1}\bra{\psi_1},\ket{\psi_1^\perp}\bra{\psi_1^\perp}\}$ in order to
certify that the prepared state is $\ket{\psi_1}$. If the state is instead
$\ket{\psi_2}$, quantum theory predicts that the outcome $\ket{\psi_1}$ is obtained
with probability $s^2$. Operationally, $s^2$ therefore quantifies the probability
of erroneously accepting $\ket{\psi_2}$ as $\ket{\psi_1}$ \cite{Schmid2018}.

For both protocols, we consider Bob's scenario which contains the preparations
\begin{equation}
\{\ket{\psi_1}\bra{\psi_1},\ket{\psi_1^\perp}\bra{\psi_1^\perp},\ket{\psi_2}\bra{\psi_2},\ket{\psi_2^\perp}\bra{\psi_2^\perp}\}
\end{equation}
The scenario always includes the corresponding projective measurements in the $\ket{\psi_1}$ and $\ket{\psi_2}$ bases, which are necessary to define the confusability. For each protocol, we augment this set of measurements with the appropriate discriminating measurement. For USD, this is given by the measurement
$\{\Pi_1^B,\Pi_2^B,\Pi_0^B\}$ of the form specified in Eqs.~\eqref{eq:pi12form} and~\eqref{eq:pi0form}, parametrized by the constant $c_1^B$ and $c_2^B$. For MESD, we instead add the measurement $\{M_1^B, M_2^B\}$ of the form in Eq.~\eqref{eq:ME-meas}, which is parametrized by $p_{1b}$ and $p_{2b}$. 

The preparations obey the operational identity
\begin{align*}
\frac{1}{2}\left(\ket{\psi_1}\bra{\psi_1}+ \ket{\psi_1^\perp}\bra{\psi_1^\perp}\right) &= \frac{1}{2}\left(\ket{\psi_2}\bra{\psi_2}+ \ket{\psi_2^\perp}\bra{\psi_2^\perp}\right) \\ &= \frac{\mathds{1}}{2}
\label{eq:q_equivalence}
\end{align*}
which gives rise to the operational equivalence
\begin{equation}
\frac{1}{2}\left(\ket{\psi_1}\bra{\psi_1}+ \ket{\psi_1^\perp}\bra{\psi_1^\perp}\right) \sim \frac{1}{2}\left(\ket{\psi_2}\bra{\psi_2}+ \ket{\psi_2^\perp}\bra{\psi_2^\perp}\right).
\label{eq:q_equivalence}
\end{equation} 
In Appendix~\ref{sec:appB}, we show that this equivalence, along with the statistics generated by projective measurements in the $\ket{\psi_1}$ and $\ket{\psi_2}$ bases, suffice to completely constrain the epistemic states in any NC model. In particular, an ontic space $\Lambda$ of size $|\Lambda|=4$ suffices, and the epistemic state matrix given by 
\begin{equation}
\label{eq:epist_state_mat}
  E = \kbordermatrix{
    & \mu_1 & \mu^\perp_1 & \mu_2 & \mu^\perp_2  \\ 
    \lambda_1 & s^2 & 0 & s^2 & 0 \\
    \lambda_2 & 1-s^2 & 0 & 0  & 1-s^2  \\
    \lambda_3 & 0 & 1-s^2 & 1-s^2 & 0  \\
    \lambda_4 & 0 & s^2 & 0 & s^2 
  } 
\end{equation} 
can be used WLOG. 

In Appendix~\ref{sec:appB} we use this epistemic state matrix to derive the maximal success rates achievable in noncontextual USD and MESD protocols. For the USD protocol, we show that this epistemic state matrix leads to the inequality
\begin{equation}
\label{eq:nn_NC}
    1-c_1^B-c_2^B\ge  0
\end{equation}
which must be obeyed by any NC representation of the measurement in Eqs.~\eqref{eq:pi12form} and~\eqref{eq:pi0form}. As a consequence, the maximal success rate $P_{\rm{USD}}^{(NC)}$ for USD protocols with an NC representation is given by
\begin{equation}
\label{eq:NC-singlebound}
    P_{\rm{USD}}^{(NC)} = \begin{cases}
    \frac{1}{2}(1-s^2), \quad &0< s\le  1 \\ 1, &s=0,
    \end{cases} 
\end{equation}
agreeing with the bound derived in Ref.~\cite{Flatt2022}.

Similarly, for MESD we show that the response function which discriminates the vectors $\mu_1$ and $\mu_2$, corresponding to the first and third columns in the matrix $E$ in Eq.~\eqref{eq:epist_state_mat}, can achieve a maximum NC success rate $P_{\rm{MESD}}^{(NC)}$ given by 
\begin{equation}
\label{eq:MESD_NCbound}
    P_{\rm{MESD}}^{(NC)} = 1-\frac{s^2}{2},
\end{equation}
agreeing with the results given in Refs.~\cite{Schmid2018} and~\cite{Flatt2022}.
In the following sections we use Eqs.~\eqref{eq:nn_NC},~\eqref{eq:NC-singlebound} and~\eqref{eq:MESD_NCbound} to determine which players in the sequential protocols possess contextual scenarios.

\section{Results}

In the previous section we analyzed Bob's scenario specifically. The noncontextual bounds apply equally to each player in the chain; only the confusability changes. To be precise, the $j$-th player's scenario instead contains the two states $\ket{\psi_1^{(j)}}\bra{\psi_1^{(j)}}, \ket{\psi_2^{(j)}}\bra{\psi_2^{(j)}}$ with confusability $s_j^2$, as well as their orthogonal partners. The corresponding projective measurements in the $\ket{\psi_1^{(j)}}$ and $\ket{\psi_2^{(j)}}$ bases are also included in their scenario, as well as the $j$-th player's optimal unambiguous or minimum-error discriminating measurement, depending on the protocol used.

\label{sec:results}

\subsection{Contextuality in sequential USD}

\begin{figure}
    \raggedright
    \includegraphics[width=0.75\columnwidth]{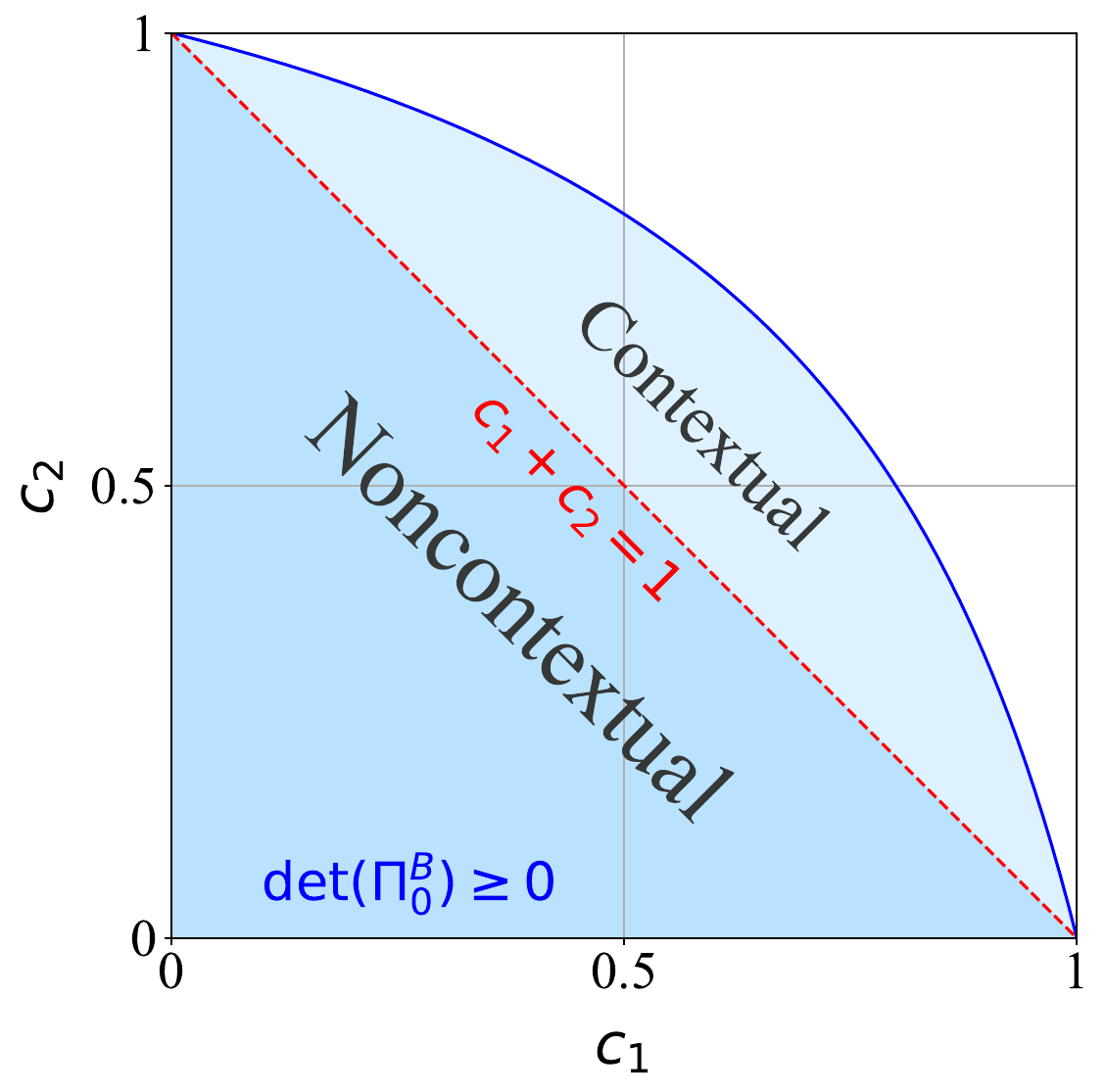}
    \caption{A visualization of the source of contextual advantage; the nonnegativity condition of the quantum model is the blue region. Here, $s=0.5$. Nonnegativity/noncontextuality of the ontological representation is bounded by the red line. For any confusability $s^2\neq0,1$ there is a margin between these boundaries, representing contextual strategies that outperform the NC bound. Only the bounding blue curve has determinant exactly 0. This means that strategies within this margin outperform the NC bound and are still viable in the sequential protocol, as the post-measurement states are distinguishable.}
    \label{fig:failure-vs-det}
\end{figure}

The origin of contextual advantage in USD can be traced directly to the different nonnegativity constraints imposed by quantum theory and by NC ontological models. In the quantum description, the optimal measurement minimizes the failure probability in Eq.~\eqref{eq:q-failure} subject to positivity of the operator $\Pi_0^B$, given in Eq.~\eqref{eq:det}. In contrast, any NC ontological model must satisfy the nonnegativity constraint \eqref{eq:nn_NC}. The latter constraint~\eqref{eq:nn_NC} is clearly more restrictive than the quantum positivity constraint~\eqref{eq:det}. This is apparent in Fig.~\ref{fig:failure-vs-det}. The blue region corresponds to allowed measurements with $\Pi_0^B\geq0$, while the red line marks the boundary of nonnegativity for NC
ontological models. This shows that quantum theory permits values of $c_1^B$ and $c_2^B$ for which $\Pi_0^B$ remains positive, while any NC representation necessarily becomes negative.

We now restrict attention to symmetric strategies with $c_1^B=c_2^B$. Such strategies are natural in communication protocols, since they do not favor one input state over another~\cite{Bergou2013}. In the symmetric case, the single-player optimal measurement simultaneously satisfies $c_1^B=c_2^B$ and $\det(\Pi_0^B)=0$, saturating the quantum nonnegativity constraint.
At the same time, this choice maximally violates the nonnegativity condition \eqref{eq:nn_NC} of the NC model. 

However, we know that the optimal single-player strategy, in fact any rank-one strategy, leaves behind post-measurement confusability $t^2=1$, so is infeasible in the multi-player case. 
In Fig.~\ref{fig:failure-vs-det}, the points on the blue boundary curve correspond to infeasible rank-one strategies. Therefore, any contextual strategy in the multi-player case must lie strictly within the contextual margin depicted. These achieve better-than-NC performance but yield nonidentical post-measurement states.

We first determine when contextual strategies arise for $n=2$. 
Charlie, as the final player, is free to choose a rank-one (zero determinant) measurement and hence to act contextually. As for Bob, we can use the fact that the optimal choice of post-measurement overlap is $t=\sqrt{s}$ to determine which values of $s$ force his strategy to be contextual. Note that in the remainder of the paper we will sometimes use the term `overlap' to describe the square-root of the confusability of two preparations. Although the overlap $s$ can be expressed as $s=\braket{\psi_1|\psi_2}$, we emphasize that it should primarily be considered simply as the square-root of the confusability, since only the confusability has an operational interpretation.
 
It is useful to quantify how strongly the noncontextuality constraint $c_1^B+c_2^B\leq 1$ can be violated by a quantum strategy. We already know that the maximal violation occurs for the symmetric choice $c_1^B=c_2^B=1/(1+s)$, for which the determinant vanishes. We therefore define
\begin{equation}
\max_{\det\Pi_0^B\geq 0}(c_1^B+c_2^B-1)
= \frac{1-s}{1+s}
=: \delta(s),
\label{eq:max_vio}
\end{equation}
which quantifies the maximum extent to which quantum theory exceeds the NC bound for a given confusability $s^2$. 

Any symmetric strategy $c_1^B=c_2^B$ employed by Bob is parametrized as
\begin{equation}
c_1^B=c_2^B=\frac{1+m\,\delta(s)}{2},
\qquad
m\in\left[-\frac{1}{\delta(s)},\,1\right],
\label{eq:c12_ansatz}
\end{equation}
where $m$ measures the degree of violation of the NC constraint. The sign of $m$ is a sufficient certificate of contextuality: $m>0$
corresponds to a contextual strategy, while $m\leq 0$ remains NC.

Requiring that Bob’s measurement leaves behind a post-measurement confusability $t^2$ imposes the consistency condition \eqref{eq:cnstcy_1}. Substituting the
ansatz \eqref{eq:c12_ansatz} into this condition yields a quadratic equation in
$m$,
\begin{equation}
m^2\delta^2(1-s^2)^2
-2m\delta(1+s^2)(1-s^2)
+(1+s^2)^2-\frac{4s^2}{t^2}=0,
\end{equation}
This equation determines whether contextuality is necessary to explain the change of confusability from $s^2$ to $t^2$. Solving this equation and substituting for $\delta(s)$ from
Eq.~\eqref{eq:max_vio}, we find that
\begin{equation}
m(s,s/t)=\frac{1}{(1-s)^2}\left(1+s^2-\frac{2s}{t}\right).
\label{eq:mval}
\end{equation}
Alternatively, this expression can be obtained by comparing the symmetric ansatz \eqref{eq:c12_ansatz} with the sequential solution
\eqref{eq:c12-branch-seq}. We keep the ratio $s/t$ as an explicit argument, as the post-measurement
confusability $t$ appears only in this form.
Note that contextuality is guaranteed when $s/t<(1+s^2)/2$, which is equivalent to the requirement that Bob's success rate, $1-s/t$, is greater than the NC bound in Eq.~\eqref{eq:NC-singlebound}.

We may immediately apply Eq.~\eqref{eq:mval} to the optimal $n$-player strategy, following Eq.~\eqref{eq:opt_n_obs}. To do so, it is useful to define the quantity $a:=s^{1/n}$, which we recall is the ratio of pre- to post-measurement overlap for \textit{every} player, i.e. $s_j/s_{j+1}=a$ for all $j$. Since the overlap of the states prepared by Alice is $s$, and each player increases this overlap by a factor $1/a$, the $j$-th player's pair of states, $\ket{\psi_1^{(j)}}$ and $\ket{\psi_2^{(j)}}$, satisfy \begin{equation}
\label{eq:overlap_chain}
\braket{\psi_1^{(j)}|\psi_2^{(j)}}=s^{(n-j+1)/n}=a^{n-j+1}.
\end{equation} Positivity of $m(a^{n-j+1},a)$ therefore certifies that the $j$-th player's PM scenario is contextual. 
We have that
\begin{equation}
    m(a^{n-j+1},a)\propto 1+a^{2(n-j+1)}-2a.
    \label{eq:m_n_obs}
\end{equation}
Note that, provided $s \notin \{0,1\}$, this function is monotonically increasing with $j$, as $a$ satisfies $0<a<1$. Therefore, contextuality propagates forward in the protocol: if the $j$-th player acts contextually, all subsequent players will too. 

Setting Eq.~\eqref{eq:m_n_obs} to 0 yields 
\begin{equation}
\label{eq:USD_first_ctxt_play}
    1+a^{2(n-j+1)}-2a=0,
\end{equation} which we solve for $j$ to determine the position of the `threshold player' who possesses the \textit{first} contextual scenario in the chain. When $a>1/2$, Eq.~\eqref{eq:USD_first_ctxt_play} admits the solution
\begin{equation}
    j_{\mathrm{USD}}=n-\frac{\ln(2a-1)}{2\ln a}+1.
    \label{eq:crit_obs_ud}
\end{equation}
That is, the $j$-th player has a contextual scenario provided that $j> j_{\mathrm{USD}}$, and $a>1/2$.

Moreover, if $j_{\mathrm{USD}}\ge n-1$, then only the final player's scenario is contextual. Solving this inequality produces a critical value $\tilde{s}_{\mathrm{USD}}(n)$, such that $s\ge\tilde{s}_{\mathrm{USD}}(n)$ guarantees noncontextuality for all players except the last. This critical value is given by 
\begin{flalign}
        \tilde{s}_{\mathrm{USD}}(n)&=\frac{1}{3^n}\left(\sqrt[3]{17+3\sqrt{33}}-\frac{2}{\sqrt[3]{17+3\sqrt{33}}}-1\right)^n  \notag \\
        &\approx 0.5437^n.
    \label{eq:crit}
\end{flalign}
Therefore, the range of values of $s$ producing a contextual scenario before the end of the chain shrinks at an exponential rate with respect to the length of the chain $n$. Consequently, in long chains, players $j<n$ enjoy a contextual advantage only for very small values of $s$.

Note that the preceding discussion required the assumption $a>1/2$, as otherwise Eq.~\eqref{eq:USD_first_ctxt_play} admits no solution; equivalently, the RHS of Eq.~\eqref{eq:crit_obs_ud} is not well-defined. In fact, when $a\le 1/2$, the contextuality witness in Eq.~\eqref{eq:m_n_obs} is clearly positive for all values of $j$. Therefore, $a\le 1/2$, or equivalently $s\le2^{-n}$, is sufficient for every player to enjoy a contextual advantage. However, it is not a necessary condition as $j_{\mathrm{USD}}$ can fall below 1 even when $a>1/2$. 

The number of contextual players in the chain, $\bar K_{\mathrm{USD}}$, is therefore given by
\begin{equation}
    \label{eq:KUSD}
    \bar K_{\mathrm{USD}}=
    \begin{cases}
        n, &a\le 1/2 \\ n-\max(0,\floor{j_{\mathrm{USD}}}).&a>1/2
    \end{cases}
\end{equation}
The expression $\max(0,\floor{j_{\mathrm{USD}}})$ is necessary because the quantity $j_{\mathrm{USD}}$ can become negative, indicating that the first contextual scenario occurs `before' the first player, i.e. every player's scenario is contextual. In the regime $a>1/2$, we define the quantity $K_\mathrm{USD}$ as the continuous analogue of $\bar{K}_{\mathrm{USD}}$, given by
\begin{equation}
\label{eq:KUSD_contin}
    K_{\mathrm{USD}}=n-j_{\mathrm{USD}}+1=\frac{\ln(2a-1)}{2\ln(a)},
\end{equation}
which approximates the number of contextual players in the chain. Note that $K_{\mathrm{USD}}$ tends to one as the length $n$ of the chain increases, as displayed in Fig.~\ref{fig:asymptotic}a. This corroborates the result in Eq.~\eqref{eq:crit}, which showed that for all but the smallest values of $s$, only the final player's scenario is contextual.

\subsection{Contextuality in sequential MESD}

The results for sequential MESD follow the same argument as those for USD. Given that the overlap $s_j$ of the states received by the $j$-th player is given by $s_j=a^{n-j+1}$, the maximum NC success rate achievable by the $j$-th player is given by $1-\frac{1}{2}a^{2(n-j+1)}$,
following Eq.~\eqref{eq:MESD_NCbound}. Furthermore, in the optimal protocol each player achieves an individual success rate of $\frac{1}{2}\left(1+\sqrt{1-a^2}\right)$, following Eq.~\eqref{eq:opt-n-obs-mesd}.
Subtracting the NC bound from the success probability achieved by the $j$-th player produces the contextuality witness
\begin{equation}
    w_j(a):=\frac{1}{2}\left(1+\sqrt{1-a^2}\right)-\left(1-\frac{a^{2(n-j+1)}}{2}\right).
    \label{eq:MESD_certifier}
\end{equation}
If this quantity is positive, the $j$-th player outperforms any NC ontological model, so their PM scenario is contextual. 

The quantity in Eq.~\eqref{eq:MESD_certifier} is monotonic in the player index $j$. Hence, if the $j$-th player's scenario is contextual, so are all subsequent players'. Solving $w_j(a)=0$ allows us to pinpoint the threshold position $j_{\mathrm{MESD}}$ after which every player's scenario is contextual. We find the solution
\begin{equation}
    j_{\mathrm{MESD}}= n-\frac{\ln\left(1-\sqrt{1-a^{2}}\right)}{2\ln a}+1
    \label{eq:crit_obs_mesd}
\end{equation}
such that $j>j_{\mathrm{MESD}}$ guarantees that the $j$-th player's scenario is contextual.

If $j_{\mathrm{MESD}}\ge n-1$, only the final player's scenario is contextual. Solving this inequality, we find a critical value of $s$, given by
\begin{equation}
    \tilde{s}_{\mathrm{MESD}}(n)=\left(\frac{\sqrt5-1}{2}\right)^{n/2} \approx 0.618^{n/2},
    \label{eq:gold_ratio}
\end{equation}
such that noncontextuality is guaranteed for all but the final player if $s\le \tilde{s}_{\mathrm{MESD}}(n)$. This implies that contextuality in sequential MESD becomes more prevalent at the end of the chain, as the length $n$ of the chain increases. In particular, the range of values of $s$ admitting no contextual scenarios before the end of the chain shrinks exponentially with the length of the chain $n$.

\begin{figure}
    \includegraphics[width=0.75\columnwidth]{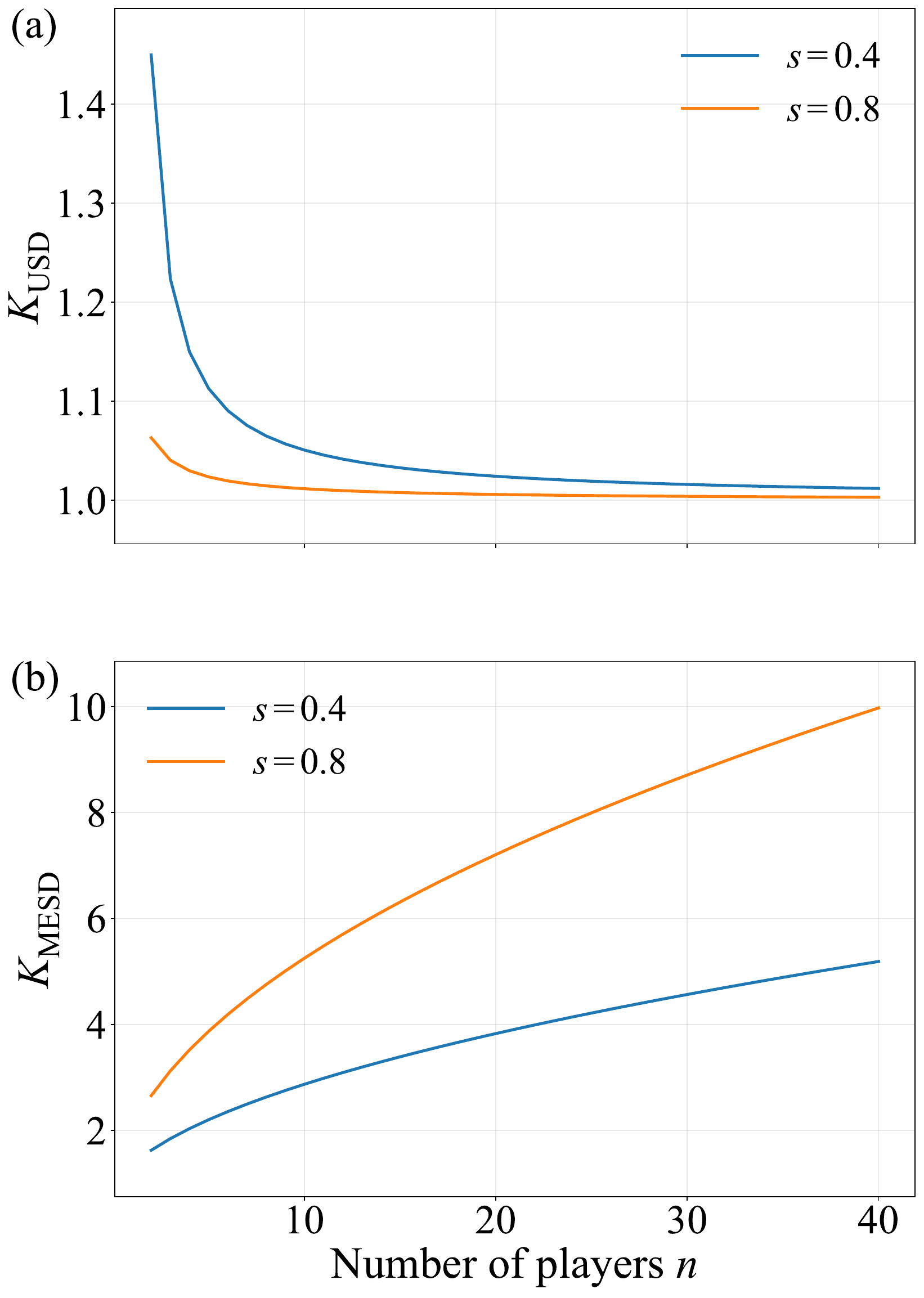}
    \caption{A comparison of the approximate number of players with a contextual scenario in each protocol, denoted by $K_{\mathrm{USD}}$ and $K_{\mathrm{MESD}}$ respectively, against number of players $n$, where $2\le n\le 40$. Note that for USD the values of $s$ are such that $a>1/2$, as is required for Eq.~\eqref{eq:KUSD_contin} to be well-defined.  For USD the number of contextual players drops rapidly, while for MESD it grows approximately at rate $\sqrt{n}$.}
    \label{fig:asymptotic}
\end{figure}

To see how the number of contextual players in the chain grows, we compute the number of contextual players in the chain, $\bar{K}_\mathrm{MESD}$, as 
\begin{equation}
    \label{eq:KMESD}
    \bar K_{\mathrm{MESD}}=
n-\max(0,\floor{j_{\mathrm{MESD}}}),
\end{equation}
analogous to Eq.~\eqref{eq:KUSD}, and introduce the quantity $K_{\mathrm{MESD}}$, given by \begin{equation}
    K_{\mathrm{MESD}}=n-j_{\mathrm{MESD}}+1=\frac{\ln(1-\sqrt{1-a^2})}{2\ln(a)},
\end{equation}
which provides a continuous approximation of $\bar{K}_\mathrm{MESD}$. 
When $n$ is large, $K_{\mathrm{MESD}}\propto \sqrt{n}$, so that the number of contextual players diverges. This is displayed in Fig.~\ref{fig:asymptotic}b. 

\subsection{Comparison of protocols}

As the number of players $n$ increases, the regime $a\le 1/2$, in which all USD players are contextual, covers a vanishingly small region of initial overlaps, requiring $s\in[0,2^{-n}]$.  In what follows we therefore focus on the regime $a>1/2$. In this regime, noncontextuality is guaranteed everywhere in the chain, except for the final player, when $s\ge\tilde{s}_{\mathrm{USD}}(n)$ and $s\le\tilde{s}_{\mathrm{MESD}}(n)$ in sequential USD and MESD respectively, where $\tilde{s}_{\mathrm{USD}}(n)\approx 0.5437^n$ and $\tilde{s}_{\mathrm{MESD}}(n)\approx 0.618^{n/2}$. This means that long chains permit multiple contextual scenarios in optimal sequential USD only if the states prepared by Alice are almost perfectly distinguishable. Conversely, in sequential MESD, a long chain admits multiple contextual scenarios as long as the states prepared by Alice are \textit{not} too easily distinguishable. Indeed, if \textit{any} player $j\neq n$ is acting contextually in one protocol, no player except the last is acting contextually in the other protocol, for fixed confusability and number of players. In this section we explain why the sequential MESD protocol is more permissive of contextual scenarios than the sequential USD protocol. We do so by focusing on a specific player close to the end of the chain, since this is where contextual scenarios can arise most easily. We show that for a player close to the end of a long chain, almost any choice of post-measurement overlap, including the optimal, will result in an NC scenario in USD. We then show that the opposite is true for MESD. 

We first note that, as the length of the chain $n$ increases, the quantity $a=s^{1/n}$ approaches one. The confusability of the states which can be received by the $j$-th player, $s_j^2$, is given by $s_j^2=a^{2(n-j+1)}$, and also approaches one. Increasing $n$ therefore pushes players close to the end of the chain into the high confusability regime. To compare the behaviors of the protocols close to the end of the chain, we consider the $j$-th player who receives states with high confusability $s_j^2$ and must leave behind states with confusability $t_j^2$ such that $s_j^2\le t_j^2\le 1$. In particular, we choose the parameterization $s_j=1-\epsilon$, with $0<\epsilon \ll 1$, and compute the range of $t_j$ values which result in contextual or NC scenarios for the $j$-th player. Note that the requirement that post-measurement states are pure is still applied in the following analysis.

For sequential USD, utilizing the contextuality witness introduced in Eq.~\eqref{eq:mval}, we have that any choice of $t_j$ such that 
\begin{equation}
\label{eq:tj_USD_cond}
    1+(1-\epsilon)^2-\frac{2(1-\epsilon)}{t_j}>0
\end{equation}
will result in a contextual scenario for the $j$-th player. We therefore define the threshold post-measurement overlap
\begin{equation}
    t_{\mathrm{USD}}^{\ast}=\frac{2(1-\epsilon)}{2(1-\epsilon)+\epsilon^2}
\end{equation}
such that $t_j>t_{\mathrm{USD}}^{\ast}$ is an equivalent condition to Eq.~\eqref{eq:tj_USD_cond}, guaranteeing a contextual scenario.
Given that $0<\epsilon \ll1$, we approximate the value of $t_{\mathrm{USD}}^{\ast}$ as follows:
\begin{equation}
\label{eq:USD_interval}
\begin{split}
    t_{\mathrm{USD}}^\ast &=1-\frac{\epsilon^2}{2(1-\epsilon)}+O(\epsilon^3)
    \\[3pt] & \approx 1-\epsilon^2/2.
    \end{split}
\end{equation}
In particular, this means that, for a contextual scenario, the post-measurement overlap $t_j$ must  lie in the $O(\epsilon^2)$-sized interval $(1-\epsilon^2/2,1]$.

We now perform the same analysis for the sequential MESD protocol, with the parameterization $s_j=1-\epsilon$ and $0<\epsilon \ll 1$. Following the argument preceding Eq.~\eqref{eq:MESD_certifier}, any post-measurement overlap $t_j$ such that
\begin{equation}
    \frac{1}{2}\left(1+\sqrt{1-\frac{(1-\epsilon)^2}{t_j^2}}\right)>1-\frac{(1-\epsilon)^2}{2}
\end{equation}
will result in a contextual scenario for the $j$-th player; the LHS corresponds to the success probability achieved by the $j$-th player when they leave behind states with post-measurement confusability $t_j^2$, and the RHS is the maximal success rate the $j$-th player can explain with an NC model. Equivalently, the scenario is guaranteed to be contextual when $t_j$ exceeds a threshold $t_{\mathrm{MESD}}^\ast$, given by
\begin{equation}
    t_{\mathrm{MESD}}^\ast=\frac{1}{\sqrt{2-(1-\epsilon)^2}}
\end{equation}
which we rewrite as
\begin{equation}
\label{eq:MESD_interval}
\begin{split}
    t_{\mathrm{MESD}}^\ast =1-\epsilon+O(\epsilon^2).
\end{split}    
\end{equation}
In particular, this means that the interval of contextual post-measurement overlaps is given by $(1-\epsilon +O(\epsilon^2),1]$. Conversely, since $s_j=1-\epsilon$, for an NC scenario the post-measurement overlap must lie in the $O(\epsilon^2)$-sized interval given by $[s_j,s_j+O(\epsilon^2)]$.

We now examine the consequences of these findings on the optimal protocol, by setting the post-measurement overlap of the $j$-th player equal to the pre-measurement overlap of the $(j+1)$-th player according to Eq.~\eqref{eq:overlap_chain}, i.e. $t_j=s_{j+1}=a^{n-j}$, where we recall $a=s^{1/n}$. Given that $s_j=1-\epsilon$, the intervals $[t_j,1]$ and $[s_j,t_j]$ have, for fixed distance $n-j$ from the end of the chain, sizes given by
\begin{equation}
\label{eq:intervals}
\begin{split}
&1-t_j=\frac{n-j}{n-j+1}\epsilon+O(\epsilon^2) \quad \text{and} \\[3pt]
&t_j-s_j=\frac{\epsilon}{n-j+1}+O(\epsilon^2)
\end{split}
\end{equation}
respectively. Note that these are both first order in $\epsilon$. However, we showed in Eq.~\eqref{eq:USD_interval} that for the $j$-th player in sequential USD to have a contextual scenario, their post-measurement overlap $t_j$ must remain within $O(\epsilon^2)$ from unity. This is therefore incompatible with the requirements of the optimal protocol given in Eq.~\eqref{eq:intervals}. Thus, for any fixed player position $j$ near the end of a long chain, the optimal USD post-measurement overlap lies outside the contextual window once $\epsilon$ is sufficiently small.

Conversely, in the discussion surrounding Eq.~\eqref{eq:MESD_interval}, we showed that for the $j$-th player in sequential MESD to have a contextual scenario, the post-measurement overlap $t_j$ must lie \textit{outside} the NC interval given by $[s_j,s_j+O(\epsilon^2)]$. In other words, the post-measurement overlap $t_j$ must exceed the pre-measurement overlap $s_j$ by an amount at least of order $\epsilon^2$. In Eq.~\eqref{eq:intervals}, we see that, in the optimal protocol, $t_j$ lies above $s_j$ by an amount of order $\epsilon$, easily clearing the NC hurdle. This explains why the sequential MESD protocol is more permissive of contextual scenarios than the sequential USD protocol.

\vspace{1em}

\section{Discussion and conclusion} \label{sec:concl}

In this paper we consider the extent to which contextuality plays a role in sequential state discrimination. We began with an overview of the optimal solutions for USD and MESD, as well as the sequential versions introduced in \cite{Bergou2013,seq-mesd}, in the case of two pure qubit states. Using the COPE formalism, we demonstrated that contextuality is inevitable not only in the optimal 2-player strategies for both protocols, but in any strategy with rank-one POVMs confined to the same plane of the Bloch sphere as the states. 

We then extended the results of Refs.~\cite{Schmid2018,Flatt2022} to the $n$-player case, to show that the requirement that each player should leave behind distinguishable post-measurement states does not preclude them from acting contextually. Our main result was a characterization of the contextuality of optimal sequential strategies, and its dependence on confusability $s^2$ and number of players $n$. This revealed a contrast between the behaviors of USD and MESD: noncontextuality was guaranteed in USD when $s>\tilde{s}_{\mathrm{USD}}(n)$, but in MESD when $s<\tilde{s}_{\mathrm{MESD}}(n)$. As $n$ increases, we found that contextuality is present only for the $n$-th player in USD, for all but the smallest values of $s$. Conversely, in MESD, the number of players with contextual scenarios grows with $n$. Finally, we characterized why sequential MESD is more permissive of contextual scenarios than sequential USD, by considering precisely which measurements result in a contextual scenario towards the end of the chain.

Contextuality is not usually considered in cases where multiple parties may act on the same system; some examples along these lines can be found in \cite{Kleinmann2011,Budroni2019}. We hope that this paper contributes to this discussion. It may also provide a useful test-bed for resource theories of contextuality, such as in \cite{Catani2024,Duarte2018,Bermejo-Vega2017}, since we can easily track the extent to which noncontextuality is violated throughout the protocol. Finally, the fact that each player performs their own measurement allowed us to consider contextuality in a separate PM scenario for each. However, these measurements can also be considered as \textit{quantum instruments}, which simultaneously model the outcome statistics of a measurement and the associated transformation from pre- to post-measurement states \cite{Davies1970}. Along these lines, contextuality in sequential USD has recently been considered in \cite{demo}. A comparison with our results would be instructive.

\section{acknowledgements}

FS gratefully acknowledges the financial support from the Engineering and Physical Sciences Research Council (EPSRC) through the Hub in Quantum Computing and Simulation grant [EP/T001062/1]. NR is grateful to be supported by the EPSRC Quantum Technologies Doctoral Training Partnership grant [EP/W524311/1].

\bibliography{bibliography}

\clearpage
\onecolumngrid
\appendix
\setlength\parindent{0pt}

\section{Obstruction to Equirank Factorization}
\label{sec:app0}
Here, we prove that any COPE matrix $C$ sharing the structure of $D$ in Eq.~\eqref{eq:COPE_struc}, with $\Rank(C)=3$, admits no ENMF. Clearly, permuting a COPE matrix has no effect on whether it admits an ENMF, so for simplicity we assume $C$ instead has the form 
\begin{center}
    \begin{equation}
        C = \vcenter{\hbox{%
        \begin{tikzpicture}
        \tikzset{square matrix/.style={
            matrix of nodes,
            column sep=-\pgflinewidth, row sep=-\pgflinewidth,
            nodes={draw,
              minimum height=#1,
              anchor=center,
              text width=#1,
              align=center,
              inner sep=0pt
            },
          },
          square matrix/.default=1.2cm
        }
        \matrix[square matrix = 0.5cm, left delimiter=(, right delimiter = )]
        {
        0 & \pnk & \pnk & \pnk & \pnk & \pnk \\
        \pnk & 0 &\pnk & \pnk & \pnk & \pnk \\
        \pnk & \pnk  & 0 & \pnk & \pnk & \pnk \\
        \pnk & \pnk & \pnk & 0 & \pnk & \pnk \\
        \pnk & \pnk & \pnk & \pnk & 0 & \pnk \\
        \pnk & \pnk & \pnk & \pnk & \pnk & 0 \\
        };
        \end{tikzpicture}}},
        \label{eq:COPE_struc2}
    \end{equation}
\end{center}
with zeroes on the diagonal. We can assume that each pair of adjacent rows in $C$ adds to $\bs{1}^\top$; both matrices considered in Sec.~\ref{sec:rank-based} can be diagonally rescaled to satisfy this constraint.  \\

We assume for the sake of contradiction that an ENMF exists:
\begin{flalign}
    C=RE, \qquad \Rank(R)=\Rank(E)=3, \qquad R,E\geq 0.
\end{flalign}
Since each pair of adjacent rows in $C$ adds to $\bs{1}^\top$, the $j$-th column $C_{:j}$ can be written as,
\begin{equation}
    C_{:j}=\begin{bmatrix}
        x_j\\1-x_j\\y_j\\1-y_j\\z_j\\1-z_j
    \end{bmatrix}
\end{equation}
so it is completely characterized by a point in the unit cube, $o_j=(x_j,y_j,z_j)\in[0,1]^3$. Since the model is NC, the same must be true of the columns $R$. Recalling that columns of $R$ correspond to specific ontic states, we write
\begin{equation}
    R_{:\lambda}= \begin{bmatrix}
        x_\lambda\\1-x_\lambda\\y_\lambda\\1-y_\lambda\\z_\lambda\\1-z_\lambda
    \end{bmatrix}
\end{equation}
so that each column is characterized by the point $q_\lambda=(x_\lambda,y_\lambda,z_\lambda)\in[0,1]^3$. The linear span of columns of $R$ is, due to the normalization of the columns, one more than the affine span of the points $q_\lambda$, i.e.
\begin{equation}
    \Rank(R)=1+\dim_{\mathrm{aff}}(q_\lambda).
\end{equation}
Since $\Rank(R)=3$, $\dim_{\mathrm{aff}}(q_\lambda)=2$, so the points $q_\lambda$ live in a plane, which we call $G$. Also, as $E$ is column stochastic, the points $o_j$ are convex combinations of $q_\lambda$, hence they also live in $G$. Since all points $o_j$ and $q_\lambda$ also live in the cube $[0,1]^3$, we wish to characterize the convex body given by the intersection $G\cap [0,1]^3$. \\

Consider that $C_{11}=0$, so that $o_1=(0,y_1,z_1)$. This implies that the first epistemic state $\mu_1(\lambda)$ has no support on any $\lambda$ satisfying $x_\lambda>0$. In other words, $C_{11}=0$ implies
$\supp(\mu_1)\subset G\cap \{x_\lambda =0\}$. Extending this argument to all diagonal elements, we find:
\begin{flalign}
   \supp(\mu_1)\subset G\cap \{x_\lambda =0\}, \qquad &\supp(\mu_2)\subset G\cap \{x_\lambda =1\} \notag \\\supp(\mu_3)\subset G\cap \{y_\lambda =0\}, \qquad &\supp(\mu_4)\subset G\cap \{y_\lambda =1\} \notag \\ \supp(\mu_5)\subset G\cap \{z_\lambda =0\},\qquad &\supp(\mu_6)\subset G\cap \{z_\lambda =1\}
   \label{eq:support_structure}
\end{flalign}

This means that, to produce the zeroes observed in the COPE, the plane $G$ must intersect all six faces of the cube, as otherwise one of the epistemic states would have no support. Hence the intersection of $G$ with the cube, which we denote
\begin{equation}
    H:=G\cap[0,1]^3,
\end{equation}
has one vertex on each of the six faces of the cube. In fact, since all off-diagonal entries of the COPE are positive, the points $o_j$ have at most one zero coordinate. Hence, the points $o_j\in H$ lie on the interior of a face of the cube. It follows that $H$ intersects the interiors of all six cube faces. Of the possible planar intersections with a cube, only a hexagon can do this, so $H$ is a hexagon. Note that, because $H$ is the intersection of $G$ and the cube $[0,1]^3$, the edges of $H$ must also lie on the boundary of the cube, so that each edge of the hexagon must lie within a face of the cube. In particular, this means each vertex of $H$ lies on an edge of the cube. \\

We denote the vertices of the hexagon as $h_j, j=1,2,\dots ,6$, and place the response functions corresponding to each $h_j$ into some nonnegative matrix $\tilde{R}$, $\Rank(\tilde{R})=3$. Since each column of $R$ is labelled by a point $q_\lambda$ which is some convex combination of these vertices, we write
\begin{equation}
    R=\tilde{R}F
\end{equation}
for some nonnegative $F$. Now, we form a new factorization where $F$ is absorbed into $E$, that is
\begin{equation}
    C=\tilde{R}\tilde{E}
\end{equation}
where $\tilde{E}=FE$. \\

Eq.~\eqref{eq:support_structure} guarantees that each epistemic state has support restricted to an edge of $H$. We recall that each edge of $H$ lies on a face of the cube, and we suppose that the edges are labelled such that $h_1$ and $h_2$ both lie on the face defined by $x=0$, such that $h_1=(0,y,z)$ and $h_2=(0,y',z')$. Then, the first column of $\tilde{E}$, $\tilde{E}_{:1}$, has support restricted to the two rows corresponding to $h_1$ and $h_2$. However, $h_1$ and $h_2$ lie on edges of the cube, so that at least one entry in each pair of non-zero coordinates $(y,z)$ and $(y',z')$ must be either 0 or 1. As the off-diagonals of $C$ are positive, each point $o_j$ therefore lies in the relative interior of the line $\{h_j,h_{j+1}\}$. Therefore, $\tilde{E}_{:1}$ must be \textit{non-zero} in the rows corresponding to \textit{both} $h_1$ and $h_2$. \\

Repeating this argument, it follows that $\tilde{E}$ must have the form
\begin{equation}
   \tilde{E} = \begin{bmatrix} a_1 & 0 & 0 & 0 & 0 & b_6 \\ b_1 & a_2 & 0 & 0 & 0 & 0 \\ 0 & b_2 & a_3 & 0 & 0 & 0 \\ 0 & 0 & b_3 & a_4 & 0 & 0 \\ 0 & 0 & 0 & b_4 & a_5 & 0 \\ 0 & 0 & 0 & 0 & b_5 & a_6 \end{bmatrix},
\end{equation}
up to permutation, in order to maintain the zeroes of $C$. Here, we have chosen to label the vertices of the hexagon $h_j$ such that $o_j$ lies in the interior of the edge $\{h_j,h_{j+1}\}$, where addition is mod 6. This is always possible, since we are free to permute columns of $C$.
Note $a_i,b_i>0$ because each point $o_j$ lies in the relative interior of its line. \\

The assumption $\Rank(E)=\Rank(C)=3$ guarantees $\Rank(\tilde{E})=3$.
However, positivity of $a_i,b_i$ guarantees $\Rank(\tilde{E})\geq 5$. To see this, suppose that the row vector $v^\top=(v_0,v_1,\dots,v_5)$ lies in the left-kernel of $\tilde E$, such that $v^\top\tilde E=0$. Then, the components of $v^\top$ satisfy 
\begin{equation}
    v_{i+1}=-\frac{a_i}{b_i}v_i \quad \text{for } i=0,1, \dots 5
\end{equation}
where addition is mod 6. If a nonzero vector solving these equations can be found, clearly it is completely determined by its first component, hence the kernel is at most one-dimensional. \\

Then $\Rank(\tilde E)\ge 5$ contradicts the assumption $\Rank(E)=3$. Therefore, a rank 3 matrix $C$ of the form~\eqref{eq:COPE_struc2} cannot simultaneously have an ENMF and zeroes on the diagonal $\square$.

\section{NC Bounds for USD and MESD}

\label{sec:appB}

Here we derive the maximum success rates for USD and MESD which admit a noncontextual model. We consider a COPE matrix $C$ whose columns correspond to the preparations $\{\ket{\psi_1}\bra{\psi_1},\ket{\psi_1^\perp}\bra{\psi_1^\perp},\ket{\psi_2}\bra{\psi_2},\ket{\psi_2^\perp}\bra{\psi_2^\perp}\}$. The first four rows of $C$ will correspond to the associated projective measurements, i.e. the four measurement-outcomes are also given by $\{\ket{\psi_1}\bra{\psi_1},\ket{\psi_1^\perp}\bra{\psi_1^\perp},\ket{\psi_2}\bra{\psi_2},\ket{\psi_2^\perp}\bra{\psi_2^\perp}\}$. For both USD and MESD we will add extra rows corresponding to unambiguous and minimum-error measurements and use the ENMF criterion in Eq.~\eqref{eq:rank} to bound their success rates. \\

We first note that the rank of $C$ must be 3, as the operational identity
\begin{align*}
\frac{1}{2}\left(\ket{\psi_1}\bra{\psi_1}+ \ket{\psi_1^\perp}\bra{\psi_1^\perp}\right) &= \frac{1}{2}\left(\ket{\psi_2}\bra{\psi_2}+ \ket{\psi_2^\perp}\bra{\psi_2^\perp}\right) \\ &= \frac{\mathds{1}}{2}
\label{eq:q_equivalence}
\end{align*}
means that only three columns of $C$ are linearly independent. Therefore, $\Rank(C)=3$ holds regardless of any measurement rows we append. \\

We now characterize our four preparations in terms of a pair of coordinates $(x,y)$. The coordinate $x$ corresponds to the probability of measurement-outcome $\ket{\psi_1}\bra{\psi_1}$ and $y$ corresponds to the probability of measurement-outcome $\ket{\psi_2}\bra{\psi_2}$. This gives rise to a submatrix
\begin{equation}
\label{eq:cxy}
C_{xy}:= \ 
    \begin{array}{c|cccc}
         &  \ket{\psi_1}\bra{\psi_1} & \ket{\psi_1^\perp}\bra{\psi_1^\perp} & \ket{\psi_2}\bra{\psi_2} & \ket{\psi_2^\perp}\bra{\psi_2^\perp}
         \\ \hline 
         x & 1 & 0 & s^2 & 1-s^2 \\ y & s^2 & 1-s^2 & 1 & 0,
    \end{array}
\end{equation}
where we define the pair $(x_j,y_j)$ as the pair of values taken by the coordinates in the $j$-th column, e.g. $(x_1,y_1)=(1,s^2)$. The rank of $C$ is 3, but due to the normalization of the columns, they lie in an affine plane of dimension $2$. Restricting to the submatrix $C_{xy}$ amounts to a projection onto the coordinates $(x,y)$. Any triplet of columns of $C_{xy}$ produces a triplet of coordinates which do \textit{not} lie on the same line, hence the columns of $C_{xy}$ span the same affine plane as those of $C$. It follows that every column $C_{:j}$ can be uniquely identified by the point $(x_j,y_j)$. 
 \\

We now suppose that $C$ admits an NC model, expressed via the ENMF $C=RE$. For any such decomposition, each column $C_{:j}$ is a  convex combination of the columns of $R$, with the weights set by the columns of $E_{:j}$, as
\begin{equation}
    C_{:j}=\sum_\lambda R_{:\lambda} E_{\lambda j}, \quad E_{\lambda j} \ge 0
\end{equation}
where 
\begin{equation}
    \quad E_{\lambda j} \ge 0, \quad \sum_{\lambda} E_{\lambda j}=1.
\end{equation}
Measurement noncontextuality demands $\Rank(C)=\Rank(R)$, so that the columns of $R$ lie in the same affine plane as those of $C$. Suppose the measurement outcomes $\ket{\psi_1}\bra{\psi_1}$ and $\ket{\psi_2}\bra{\psi_2}$ are represented by the response functions $\xi_1(\lambda)$ and $\xi_2(\lambda)$, which are also rows of the matrix $R$. Each column of $R$ corresponds to a specific ontic point, so we label these columns as $R_{:\lambda}$. We now project these onto the rows labelled by $\xi_1$ and $\xi_2$, just as we projected $C$ onto the corresponding measurement outcomes to produce $C_{xy}$ in Eq.~\eqref{eq:cxy}, and we label the two resulting values as $(x_\lambda,y_\lambda)$. This allows us to write
\begin{equation}
\label{eq:conv_comb_pts}
    (x_j,y_j)=\sum_{\lambda}E_{\lambda j}(x_\lambda,y_\lambda),
\end{equation}
which simply means that the points $(x_j,y_j)$ arise as convex combinations of points $(x_\lambda,y_\lambda)$. Recalling that the columns of $C$ and $R$ live in the same affine plane and that the points $\{(x_j,y_j)\}_{j=1}^4$ span this plane, it follows that the set of points $\{(x_\lambda,y_\lambda)\}_{\lambda}$ must also span this plane, and so any column $R_{\lambda}$ can be uniquely specified by
the point $(x_\lambda,y_\lambda)$. As all entries are probability-valued, we also have $(x_j,y_j)\in[0,1]$ and $(x_\lambda,y_\lambda)\in[0,1]$. \\

We note that the first column $C_{:1}$ has coordinates $(x_1,y_1)=(1,s^2)$, and so lies in the interior of the edge between $(1,0)$ and $(1,1)$. Considering Eq.~\eqref{eq:conv_comb_pts}, it follows that $E_{\lambda 1}$ may only have support on points $\lambda$ such that $x_\lambda =1$.
Extending this analysis to the other columns, we see that
\begin{equation}
\label{eq:sd_supp}
    \begin{aligned}
        E_{\lambda 1}>0 \implies x_\lambda = 1 \\
        E_{\lambda 2}>0 \implies x_\lambda = 0 \\
        E_{\lambda 3}>0 \implies y_\lambda = 1 \\
        E_{\lambda 4}>0 \implies y_\lambda = 0
    \end{aligned}
\end{equation}
must hold. This implies that epistemic states such as $\mu_1$ and $\mu_1^\perp$, corresponding to perfectly distinguishable preparations, share no support, which is to be expected. \\

Preparation noncontextuality applied to the equivalence Eq.~\eqref{eq:q_equivalence} requires that the columns of $E$ satisfy
\begin{equation}
\label{eq:col_mix}
    \frac{1}{2}(E_{:1}+E_{:2})=\frac{1}{2}(E_{:3}+E_{:4}).
\end{equation}
Now suppose a particular ontic point $\lambda$ is such that $(x_\lambda,y_\lambda)=(1,a)$ where $(0<a<1)$. According to Eq.~\eqref{eq:sd_supp}, such an ontic point cannot have nonzero support in either column $E_{:3}$ or $E_{:4}$. However, Eq.~\eqref{eq:col_mix} then implies that it also cannot have nonzero support in $E_{:1}$ or $E_{:2}$. Extending this argument, it follows that the only ontic points which can have nonzero support in $E$ must have coordinates in $\{0,1\}^2$. We therefore define these ontic points as
\begin{equation}
    \lambda_1=(1,1),\lambda_2=(1,0),\lambda_3=(0,1),\lambda_4=(0,0),
\end{equation}
as any other ontic point (column of $R$) is necessarily unused. After discarding unused columns and merging identical columns in $R$, it must have the form 
\begin{equation}
\label{eq:resp_mat}
  R = \kbordermatrix{
    & \lambda_1 & \lambda_2 & \lambda_3 & \lambda_4  \\[6pt]
    \xi_1 & 1 & 1 & 0 & 0 \\[4pt]
    \xi_1^\perp & 0 & 0 & 1  & 1  \\[4pt]
    \xi_2 & 1 & 0 & 1 & 0  \\[4pt]
    \xi_2^\perp  & 0 & 1 & 0 & 1 
  } 
\end{equation} \\

from which it follows that the epistemic state matrix has the form \\
\begin{equation}
\label{eq:epist_state_mat}
  E = \kbordermatrix{
    & \mu_1 & \mu^\perp_1 & \mu_2 & \mu^\perp_2  \\ [6pt]
    \lambda_1 & s^2 & 0 & s^2 & 0 \\[4pt]
    \lambda_2 & 1-s^2 & 0 & 0  & 1-s^2  \\[4pt]
    \lambda_3 & 0 & 1-s^2 & 1-s^2 & 0  \\[4pt]
    \lambda_4 & 0 & s^2 & 0 & s^2 
  } . 
\end{equation} \\ 
The preceding arguments guarantee that this epistemic state matrix can be used WLOG, regardless of the discriminating measurements subsequently added to the COPE matrix $C$.  \\

All that remains is to find noncontextual representations of our USD and MESD measurements which achieve the maximal success rates and remain nonnegative and normalized. The measurement outcomes
\begin{equation}
\Pi_1^B = c_1^B\ket{\psi_2^\perp}\bra{\psi_2^\perp}, \quad \Pi_2^B = c_2^B\ket{\psi_1^\perp}\bra{\psi_1^\perp}
\end{equation}
clearly must be represented by the response functions $c_1^B\xi_1^\perp$ and $c_2^B\xi_2^\perp$ respectively in order for measurement noncontextuality to hold. The success probability under equal priors therefore obeys the bound
\begin{equation}
\begin{aligned}
P^{(NC)}_{\rm{USD}}&=
    \frac{1}{2}\langle c_2^B\xi_1^\perp,\mu_2\rangle+\frac{1}{2}\langle c_1^B\xi_2^\perp,\mu_1\rangle \\ &= \frac{1-s^2}{2}\left(c_1^B+c_2^B\right) .
    \end{aligned}
\end{equation}
Note that $c_1^B\xi_2^\perp (\lambda_4)=c_1^B$ and $c_2^B\xi_1^\perp (\lambda_4)=c_2^B$. Due to normalization of measurements the response function $\xi_0$, representing the inconclusive outcome $\Pi_0^B$, must satisfy $\xi_0(\lambda_4)=1-c_1^B-c_2^B$. For nonnegativity we must have that $c_1^B+c_2^B\le 1$ and therefore the success probability obeys the bound
\begin{equation}
    P_{\rm{USD}}^{(NC)}\le \frac{1-s^2}{2}.
\end{equation}
Note that this holds for $s>0$; when $s=0$, perfect discrimination is clearly possible and so the noncontextual success rate is 1.\\

Exactly the same argument can be applied to MESD to restrict the epistemic state matrix to the form given in Eq.~\eqref{eq:epist_state_mat}. If we represent the minimum-error discriminating measurement by the response functions $\xi_1^{\rm{ME}}$ and $\bs{1}^\top - \xi_1^{\rm{ME}}$, the success rate obeys the bound
\begin{equation}
\begin{aligned}
P^{(NC)}_{\rm{MESD}}&=
    \frac{1}{2}(p_{b1}+p_{b2}) \\ &=\frac{1}{2}(\langle \xi_1^{\rm{ME}}, \mu_1\rangle +\langle  \bs{1}^\top-\xi_1^{\rm{ME}},\mu_2\rangle) \\ & \le 1-\frac{s^2}{2}
\end{aligned}
\end{equation}
where the optimal response function satisfies either $\xi_1^{\rm{ME}}(\lambda_2)=1$ and $\xi_1^{\rm{ME}}(\lambda_3)=0$ or vice versa.